\pdfoutput=1 

\documentclass[preprint,journal]{vgtc}            %

\preprinttext{To appear in IEEE Transactions on Visualization and Computer Graphics.}

\ieeedoi{xx.xxxx/TVCG.201x.xxxxxxx}

\title{From Information to Choice: A Critical Inquiry Into\\Visualization Tools for Decision Making 
}

\author{%
  \authororcid{Emre Oral}{0000-0001-9924-2576} 
  Ria Chawla, 
  Michel Wijkstra, 
  Narges Mahyar,
  and 
  \authororcid{Evanthia Dimara}{0000-0001-5212-7888} 
}

\authorfooter{
  \item
  	E. Oral, M Wijkstra and E.Dimara are with Utrecht University, Netherlands
    Email: e.oral@uu.nl 

  \item
  	R. Chawl and N. Mahyar are with University of Massachusetts Amherst, United States
}

\usepackage{colortbl}
\usepackage{xspace}
\usepackage{color}
\usepackage{longtable,rotating}			%
\usepackage{supertabular}
\usepackage{multirow}
\usepackage{marginnote}
\usepackage{rotating}
\usepackage{multicol}
\usepackage{caption}
\usepackage{wrapfig}
\usepackage{float}
\usepackage{xltabular}
\usepackage{booktabs}
\usepackage{url}
\usepackage{enumitem}

\abstract{%
In the face of complex decisions, people often engage in a three-stage process that spans from (1) exploring and analyzing pertinent information (intelligence); (2) generating and exploring alternative options (design); and ultimately culminating in (3) selecting the optimal decision by evaluating discerning criteria (choice). We can fairly assume that all good visualizations aid in the “intelligence” stage by enabling data exploration and analysis. Yet, to what degree and how do visualization systems currently support the other decision making stages, namely “design” and “choice”? To further explore this question, we conducted a comprehensive review of decision-focused visualization tools by examining publications in major visualization journals and conferences, including VIS, EuroVis, and CHI, spanning all available years. We employed a deductive coding method and in-depth analysis to assess whether and how visualization tools support design and choice. Specifically, we examined each visualization tool by (i) its degree of visibility for displaying decision alternatives, criteria, and preferences, and (ii) its degree of flexibility for offering means to manipulate the decision alternatives, criteria, and preferences with interactions such as adding, modifying, changing mapping, and filtering. {\color{black} 
Our review highlights the opportunities and challenges that decision-focused visualization tools face in realizing their full potential to support all stages of the decision making process. It reveals a surprising scarcity of tools that support all stages, and while most tools excel in offering visibility for decision criteria and alternatives, the degree of flexibility to manipulate these elements is often limited, and the lack of tools that accommodate decision preferences and their elicitation is notable. Based on our findings, to better support the choice stage, future research could explore enhancing flexibility levels and variety, exploring novel visualization paradigms, increasing algorithmic support, and ensuring that this automation is user-controlled via the enhanced flexibility levels.
} 
 Our curated list of the {\color{black}88} surveyed visualization tools is available in the OSF link {\color{black}(\url{https://osf.io/nrasz/?view_only=b92a90a34ae241449b5f2cd33383bfcb})}.
}


\graphicspath{{figs/}{figures/}
{./}} %

\begin{document}

\maketitle

\definecolor{domainyes}{HTML}{FFFFFF}   %
\definecolor{domainno}{HTML}{4d4d4d}    %

\definecolor{flexcolor3}{HTML}{B12024}  %
\definecolor{flexcolor2}{HTML}{C86266}  %
\definecolor{flexcolor1}{HTML}{E1A6A8}  %
\definecolor{flexcolor0}{HTML}{FFFFFF}  %

\definecolor{viscolor2}{HTML}{28749D}   %
\definecolor{viscolor1}{HTML}{80ADC5}   %
\definecolor{viscolor0}{HTML}{FFFFFF}   %

\definecolor{exprcolor5}{HTML}{97509F}  %
\definecolor{exprcolor4}{HTML}{ab72b2}  %
\definecolor{exprcolor3}{HTML}{c096c5}  %
\definecolor{exprcolor2}{HTML}{d5b9d8}  %
\definecolor{exprcolor1}{HTML}{eadceb}  %
\definecolor{exprcolor0}{HTML}{FFFFFF}  %

\newcommand{\evanthia}[1]{\textcolor{magenta}{E:#1}}
\newcommand{\evanthialeft}[1]{\textcolor{magenta}{\small{[ED]\marginnote{[ED: #1]}}}}
\newcommand{\evanthiaright}[1]{\textcolor{magenta}{\scriptsize{[ED]\reversemarginpar{\marginnote{[ED: #1]}}}}}
\newcommand{\emre}[1]{\textcolor{olive}{EO:#1}}

\newcommand{\revision}[1]{%
    \vspace{-0.5pt}%
    \textcolor{black}{#1}%
    \vspace{-0.5pt}%
}

\newcommand{\narges}[1]{\textcolor{blue}{NM:#1}}
\newcommand{\needsimprovement}[1]{\textcolor{teal}{#1}}
\newcommand{\trackchanges}[1]{\textcolor{black}{#1}}
\newcommand{\highlightme}[1]{\textcolor{black}{#1}}

\newcommand{\noOfCorpus}{88\xspace}
\newcommand{\noOfCHOICETools}{27\xspace}
\newcommand{\noOfNoDecisionClaimTools}{28\xspace}
\newcommand{\noOfDecisionClaimTools}{33\xspace}
\newcommand{\noOfMCDMTools}{19\xspace}
\newcommand{\noOfDRTools}{2\xspace}
\newcommand{\noOfMLTools}{2\xspace}

\newcommand{\tag}[1]{\textsc{#1}\xspace}

\newcommand{\noOfVisTools}{88\xspace}
\newcommand{\noOfNoFlexible}{26\xspace}
\newcommand{\noOfLowFlexible}{28\xspace}
\newcommand{\noOfMediumFlexible}{27\xspace}
\newcommand{\noOfHighFlexible}{7\xspace}
\newcommand{\noOfNoVisible}{37\xspace}
\newcommand{\noOfLowVisible}{33\xspace}
\newcommand{\noOfHighVisible}{18\xspace}
\newcommand{\noOfNoExpressive}{19\xspace}
\newcommand{\noOfLowExpressive}{13\xspace}
\newcommand{\noOfMediumToLowExpressive}{23\xspace}
\newcommand{\noOfMediumExpressive}{21\xspace}
\newcommand{\noOfMediumToHighExpressive}{10\xspace}
\newcommand{\noOfHighExpressive}{2\xspace}
\newcommand{\noOfGeneric}{35\xspace}
\newcommand{\noOfSpecific}{53\xspace}

\newcommand{\noOfDecisionVisPapers}{131\xspace}
\newcommand{\visibleScoreTwoToolsPercent}{20\%\xspace}
\newcommand{\visibleScoreOneToolsPercent}{38\%\xspace}
\newcommand{\visibleScoreZeroToolsPercent}{42\%\xspace}
\newcommand{\visibleScoreAtLeastOneToolsPercent}{58\%\xspace}
\newcommand{\flexibleScoreThreeToolsPercent}{8\%\xspace}
\newcommand{\flexibleScoreTwoToolsPercent}{31\%\xspace}
\newcommand{\flexibleScoreOneToolsPercent}{32\%\xspace}
\newcommand{\flexibleScoreZeroToolsPercent}{30\%\xspace}
\newcommand{\flexibleScoreAtLeastOneToolsPercent}{70\%\xspace}
\newcommand{\expressiveMaxToolsPercent}{2\%\xspace}
\newcommand{\expressiveFourToolsPercent}{11\%\xspace}
\newcommand{\expressiveThreeToolsPercent}{24\%\xspace}
\newcommand{\expressiveTwoToolsPercent}{26\%\xspace}
\newcommand{\expressiveOneToolsPercent}{15\%\xspace}
\newcommand{\expressiveZeroToolsPercent}{22\%\xspace}
\newcommand{\generalizabilityYesToolsPercent}{40\%\xspace}
\newcommand{\generalizabilityNoToolsPercent}{60\%\xspace}

\newcommand{\ourParskip}{\vspace{0.3\baselineskip}}
\newcommand{\bfheader}[1]{\ourParskip\noindent\textbf{#1}:}
\newcommand{\quotes}[1]{``\textit{#1}''\xspace}%

\newcommand{\term}[1]{\textsc{#1}}

\newcommand{\flexibility}{\term{flexibility}\xspace}
\newcommand{\visibility}{\term{visibility}\xspace}
\newcommand{\expressiveness}{\term{expressiveness}\xspace}
\newcommand{\generalizability}{\term{generalizability}\xspace}

\newcommand{\flexibilityNotion}{flexibility\xspace}
\newcommand{\visibilityNotion}{visibility\xspace}
\newcommand{\expressivenessNotion}{expressiveness\xspace}
\newcommand{\generalizabilityNotion}{generalizability\xspace}

\newcounter{magicrownumbers2}
\newcommand\visdssrownumber{\stepcounter{magicrownumbers2}\arabic{magicrownumbers2}}

\newcommand{\catDomainNo}[1]{ {\hypersetup{citecolor=white}\cellcolor{domainno}\color{domainno}1 #1}}
\newcommand{\catDomainYes}[1]{ {\hypersetup{citecolor=white}\cellcolor{domainyes}\color{domainyes}0 #1}}

\newcommand{\catVisibilityHigh}[1]{ 
{\hypersetup{citecolor=white}\cellcolor{viscolor2}\color{viscolor2}2 #1}}
\newcommand{\catVisibilityLow}[1]{ {\hypersetup{citecolor=white}\cellcolor{viscolor1}\color{viscolor1}1 #1}}
\newcommand{\catNoVisibility}[1]{ {\hypersetup{citecolor=white}\cellcolor{viscolor0}\color{viscolor0}0 #1}}

\newcommand{\catFlexibilityHigh}[1]{ {\hypersetup{citecolor=white}\cellcolor{flexcolor3}\color{flexcolor3}3 #1}}
\newcommand{\catFlexibilityMedium}[1]{ {\hypersetup{citecolor=white}\cellcolor{flexcolor2}\color{flexcolor2}2 #1}}
\newcommand{\catFlexibilityLow}[1]{ {\hypersetup{citecolor=white}\cellcolor{flexcolor1}\color{flexcolor1}1 #1}}
\newcommand{\catNoFlexibility}[1]{ {\hypersetup{citecolor=white}\cellcolor{flexcolor0}\color{flexcolor0}0 #1}}

\newcommand{\catExpressivenessHigh}[1]{ {\hypersetup{citecolor=white,}\cellcolor{exprcolor5}\color{exprcolor5}5 #1}}

\newcommand{\catExpressivenessMediumToHigh}[1]{ {\hypersetup{citecolor=white}\cellcolor{exprcolor4}\color{exprcolor4}4 #1}}
\newcommand{\catExpressivenessMedium}[1]{ {\hypersetup{citecolor=white}\cellcolor{exprcolor3}\color{exprcolor3}3 #1}}
\newcommand{\catExpressivenessMediumToLow}[1]{ {\hypersetup{citecolor=white}\cellcolor{exprcolor2}\color{exprcolor2}2 #1}}
\newcommand{\catExpressivenessLow}[1]{ {\hypersetup{citecolor=white}\cellcolor{exprcolor1}\color{exprcolor1}1 #1}}
\newcommand{\catNoExpressiveness}[1]{ {\hypersetup{citecolor=white}\cellcolor{exprcolor0}\color{exprcolor0}0 #1}}

\newcommand\crule[3][black]{\textcolor{#1}{\rule{#2}{#3}}}
\newcommand{\textrect }[1]{\crule[#1]{0.3cm}{0.3cm}}

\section{Introduction}
\label{sec:introduction}

Information understanding can take various forms depending on the end goal. Consider a city planner making critical decisions about urban development and a researcher analyzing the broader impacts of such projects. Both consider environmental, social, and economic factors, but their approaches differ. The city planner’s workflow may direct focus to key factors, breaking down complex information, synthesizing alternative plans, weighing pros and cons, and combining data to \emph{choose} the optimal plan. In contrast, the researcher may embark on a more exploratory journey, 
diving deep into various  other aspects of urban development and analyzing them in broader contexts. 
They may synthesize findings, identify trends, and draw connections between components, without the constraints of a specific decision making goal. 

Given the distinct needs and objectives of decision makers and other analysts   \cite{Dimara2021Organizations,mahyar2019civic}, 
 they would likely benefit from different visualization systems tailored to their unique information processing requirements.
 A city planner may require a visualization system emphasizing key decision criteria and preferences with interactive capabilities for iterating over decision alternatives and adapting to decision making needs \cite{Dhurkari2022Mcdm,Zavadskas2011MCDM}.
In contrast, an analyst might prefer a system enabling extensive data exploration and the discovery of connections and patterns without being constrained or guided by specific criteria or goals \cite{Keefe2010Workflows,stolte2002polaris}. 
This paper investigates how visualization design can evolve to accommodate the distinct needs of decision making.

Decision making has undoubtedly been studied in the context of visualization, 
including choice experiments based on one \cite{Castro2022Effort}, two \cite{Zhang2015Remediating, Dimara2019Mitigating} or multiple criteria \cite{Dimara2018DecisionSupport,zhao2018evaluating}, and visualizations proposed to assist decision making for general purposes      \cite{Carenini2004ValueCharts,Pajer2017Weightlifter,Gratzl2013Lineup,Weng2018Homefinder}  or within 
applications such as 
engineering 
\cite{Cibulski2020PAVED}, urban planning \cite{Feng2020Topology},  advertisement \cite{Liu2016Smartadp}, or sport management \cite{Cao2022TeamBuilder}. However, recent findings  suggest that visualization research may not yet fully support real-world decision making \cite{Dimara2021Organizations,Dimara2021DecisionTasks,tory2021finding,brehmer2021jam,alves2023exploring,kosara2023notebooks}. Evidence indicates that professional decision makers are an under-supported community by visualization \cite{Dimara2021Organizations,tory2021finding,brehmer2021jam}. An analysis of 940 commercial tools reported by the participants revealed that there is no "decision making" visualization tool that they can use during their decision making workflow \cite{Dimara2021Organizations}. Meanwhile, decision makers report a desire to improve their understanding of the data involved in their decisions since coordinating with the data analysis team is not always effective \cite{Dimara2021Organizations, han2022kicking,Kandogan2014Insight}. Notably, neither the dedicated decision support software reviews  \cite{Weistroffer2016MCDASoftware, Greco2016MCDA}  report visualization design as an inherent component of their functionality (although they inevitably do contain some visual representations). In addition, research on more complex visual analytic systems reports on a similar trend. Two recent visualization surveys in the context of explainable AI \cite{Chatzimparmpas2020Trust,Sperrle2021MLEvaluations} revealed that visualization design focuses on displaying the inner workings of models, but not on how end-users can involve the model in their decision making process. However, it remains unclear whether this issue is inherent to AI models or also applies to all visualization tools designed for decision support. 
Furthermore, the extent to which the primary design goal of all aforementioned systems was explicitly aimed at facilitating human decision making, as well as the specific design features employed to support the decision making process, is still uncertain.

Additionally, another recent survey \cite{Dimara2021DecisionTasks} revealed that  visualization tools have not been evaluated based on their effectiveness in supporting decision making tasks.  This methodological challenge has been attributed to the lack of foundations and guidance from decision theory in  the visualization literature\cite{Dimara2021DecisionTasks}. 
However, this survey  \cite{Dimara2021DecisionTasks} focused solely on the choice of evaluation tasks by the researchers and not on whether and how visualization designs support decision making.

While evidence suggests a misalignment between decision making goals and potential visualization support, no survey paper has examined to what degree and how visualization design aids decision making. 
Conducting such a survey entails overcoming certain challenges, particularly the need for a reliable method to identify 'decision-focused' visualization tools. 
Establishing this distinction is crucial, as a lack of clear criteria could lead to the oversimplified assumption that all effective visual analytic systems inherently support decisions.
Moreover, even after deriving a list of decision-focused visualization tools, it remains unclear which criteria, relevant to decision making, should be used for their systematic assessment.
 Addressing both questions requires a deeper understanding of decision making and its components that can be aided by visualization.
Our paper grounds its investigation on the foundations of decision sciences, evaluating the extent to which visualizations support specific aspects of the decision making process.

To examine the extent and manner in which current visualization tools support decision making, \revision{we review 88 papers mentioning the word \quotes{decision} in their titles or abstracts,}
from leading visualization venues like VIS, EuroVis,
and CHI, covering all available years. 
Drawing from the literature on decision making and visualization, we devise an approach to identify "decision-focused" visualization tools by relying on established stages of decision making, namely \tag{intelligence}, \tag{design}, and \tag{choice}, along with their embedded properties.  Employing a deductive coding method, we begin with preconceived metrics derived from the literature that are associated with properties of interactive visualization design relevant to decision making. Specifically, we assess each visualization tool based on (i) its degree of \tag{visibility}, which pertains to the display of alternative options, criteria, and preferences, and (ii) its degree of \tag{flexibility}, which involves offering means to manipulate alternative options, criteria, and preferences through interactions such as adding, modifying, changing mappings, and filtering. We ultimately explore and discuss the opportunities and challenges faced by visualization tools in realizing their full potential to support every stage of the decision making process.

\section{Background: Decision Metrics \& Tags Rationale }
\label{sec:metrics_tags}

This section discusses literature from decision sciences and visualization to provide
a deeper understanding of what decision making is and why and how it can be potentially aided by data visualization. 

\subsection{What is Decision Making?}

In the introductory example, 
we outlined  differences between decision makers and other analysts on their information processing workflow \cite{reyna2011dual}.
Such differences have been identified in the literature  for
 decision making, 
including  \emph{attention}, \emph{simplification}, \emph{evaluation}, and  \emph{integration}.

Regarding \emph{attention}, when processing information for decision making, attention must often be  directed toward key factors relevant to the decision, filtering out less pertinent information \cite{payne1988adaptive,payne2004walking}.  
\highlightme{In our example, the decision maker may need a visualization tool that emphasizes  factors critical to urban development, like traffic patterns and environmental regulations, and zoning restrictions.  In contrast, the analyst may 
 need to delve into a broader array of factors, investigating social demographics, economic impacts, and architectural considerations, all without the constraint of  prioritizing certain factors over others.}

In \emph{simplification}, decision makers often simplify complex information by breaking it down into manageable chunks, categorizing data, and identifying patterns or themes \cite{russo1983strategies,payne2004walking}.
This helps to reduce the cognitive load and make the decision process more efficient \cite{russo1983strategies,payne2004walking}. 
In contrast, the analyst may not appreciate prompts for the simplification of complex information into manageable chunks from a visualization tool in the same way. 
They might need tools to probe data complexity, analyze thoroughly, and reveal hidden relationships.

In \emph{evaluation}, decision makers evaluate and compare information relevant to the decision against other alternatives to determine its value or importance \cite{keeney1982decision}. This involves weighing the pros and cons of various options, assessing the likelihood of outcomes, and considering the risks and benefits of each choice \cite{keeney1982decision}.
In comparison, the evaluation process for the analyst may be less focused on weighing the costs and benefits of specific alternatives, and more on assessing the quality of the data, the robustness of the analysis methodologies, and the overall significance of their findings for the broader field of urban development.

 In \emph{integration}, decision makers integrate the information they gathered and evaluated into a coherent, actionable plan \cite{ham2009integrating,greening2004design}. 
 The city planner may need a visualization tool that helps identify trade-offs, reconciles conflicting information, and makes a final decision that balances competing demands and priorities. For the analyst, integration may be less about reconciling conflicting information to make a final decision and more about synthesizing findings to generate new insights, identify trends, and contribute to the existing body of knowledge.

We previously contrasted decision making with other data analysis types.
 Now, we discuss Simon's three stages of decision making \cite{Simon1960ManagementDecision}— (1)intelligence, (2) design, and (3) choice—to help further identify distinct characteristics of decision making tasks and guide the development of visualization tools tailored to support decision makers effectively.

\subsubsection{Decision making stage-1: INTELLIGENCE}
\label{sec:metrics_tags_intelligence}

In the \tag{intelligence} stage of Simon's model \cite{Simon1960ManagementDecision}, 
the decision maker explores and analyzes pertinent information. This stage is closely related to the concept of \emph{attention}, with the decision maker focusing  on key factors relevant to the decision,  filtering out irrelevant or distracting information, and directing attention towards key data sources.

\subsubsection{Decision making stage-2: DESIGN}
\label{sec:metrics_tags_design}

In the \tag{design} stage of Simon's model\cite{Simon1960ManagementDecision}, the decision maker generates and explores alternative options. This stage requires \emph{simplification} of the previous stage, breaking complex information down into manageable chunks or categories. Formal methods for the \tag{design} stage can be found in the field of Decision Analysis, which draws on decision theory, operations research, statistics, and behavioral sciences to provide a structured approach to making decisions in complex and uncertain situations, developing decision models and representations of the problem, and identifying decision criteria and alternatives \cite{keeney1982decision}.
 \highlightme{Yet, while visualization systems incorporating formal methods can be effective, formal methods may not fully support all the creative activities of the \tag{design} stage that a decision maker engages in, such as brainstorming, scenario planning, mental simulations, and intuitive judgments, all of which play a role in generating and exploring alternative options \cite{chermack2011scenario, klein2008naturalistic}.}

\subsubsection{Decision making stage-3: CHOICE}
\label{sec:metrics_tags_choice}

In the \tag{choice} stage of Simon's model\cite{Simon1960ManagementDecision},  the decision maker selects the optimal decision by evaluating discerning criteria.
This stage is closely related to the \emph{evaluation} of options generated in the previous stage, including considerations of weights and preferences,  costs and benefits, and potential outcomes of each option, as well  as likelihood assessments of success. 
 A level of \emph{integration} occurs in all three stages of Simon's model but is especially true during  \tag{choice},  
 where the decision maker integrates the information they have gathered and evaluated into a coherent plan or decision. 
 They may need to reconcile conflicts, prioritize factors, integrate various data types, and consider decision context.

\tag{choice} stage studies involving a small set of criteria, such as risk or profit, are common in visualization \cite{Savikhin2011FinancialPortfolio,Kale2020VisualReasoning,Zhang2015Remediating}. 
The \tag{choice} stage with many criteria is known in visualization literature as a \quotes{multi-attribute choice task} \cite{Dimara2018DecisionSupport} or \quotes{preferential choice}, introduced in the context of multi-criteria decision making visualization tools, such as Value Charts \cite{Carenini2004ValueCharts, Bautista2006Framework}, Attribute Explorer \cite{Spence1998AttributeExplorer}, SmartClient \cite{Pu2000SmartClient} and EZChooser \cite{Wittenburg2001ParallelBargrams}. 
Multi-criteria decision making (\tag{mcdm}) is a sub-discipline of operations research that builds upon decision theory \cite{Keeney1993Decisions} and formalizes and develops methods to aid multi-attribute choice tasks \cite{Zavadskas2011MCDM}. 
\tag{mcdm} involves the decision maker's judgment \cite{Dhurkari2022Mcdm} as input for various \tag{mcdm} methods for solving decision problems \cite{Velasquez2013Analysis}. Examples of \tag{mcdm} methods include PROMETHEE, which focuses on ordering criteria, MAUT, which requires extensive input values from the decision maker, or SAW, which needs precise weighting operations in each step \cite{Velasquez2013Analysis}.
According to the \tag{mcdm} formalizations,  in the \tag{choice} stage of decision making, the following three elements are considered:
\begin{itemize}
    \item Decision \emph{alternatives}: the options to address a specific problem or goal. For the city planner, the alternatives might include building a new highway or improving public transportation.
    \item Decision \emph{criteria}: the factors used to evaluate and compare the alternatives. For the city planner, 
     the decision criteria could involve environmental impact, cost, and traffic congestion.
    \item Decision  \emph{preferences} or weights: the importance assigned to each criterion based on the decision maker's priorities. For the city planner, the preferences would depend on their emphasis on sustainability, budget constraints, or reducing traffic bottlenecks. 
\end{itemize}

\subsubsection{Conclusion for 'decision-focused' visualizations}
\label{sec:metrics_tags_decisionfocusedvis}
\vspace{-0.25em}

Our previous discussions revolved around decision making stages and properties. \revision{To identify \quotes{decision-focused} visualization tools, we follow the requirement proposed by Dimara \& Stasko \cite{Dimara2021DecisionTasks}: a decision-focused visualization tool should at minimum support the \tag{choice} stage, visualizing some decision alternatives. Inclusion of  \tag{intelligence} and \tag{design} stages can be optional but advantageous.} 
For the \tag{choice} stage, \tag{mcdm}  methods and properties share a common design requirement:  flexible iterations for comparisons \cite{Velasquez2013Analysis}, supported by transparency  to establish the \quotes{one-to-one correspondence} between criteria values and the evaluation of decision alternatives \cite{Velasquez2013Analysis}.
\revision{
Yet, it is essential to acknowledge that excluding tools lacking a \tag{choice} stage may not fully encompass the value of visualization in other decision stages. For instance, medical diagnostic tools play a significant role in the \tag{intelligence} stage by aiding in the identification of problems and specifying symptoms. Therefore, future research should explore the role of visualizations across all stages and types of decision making.}

\vspace{-1em}
\begin{figure}[H]
 \centering 
 \includegraphics[width=0.95\columnwidth]{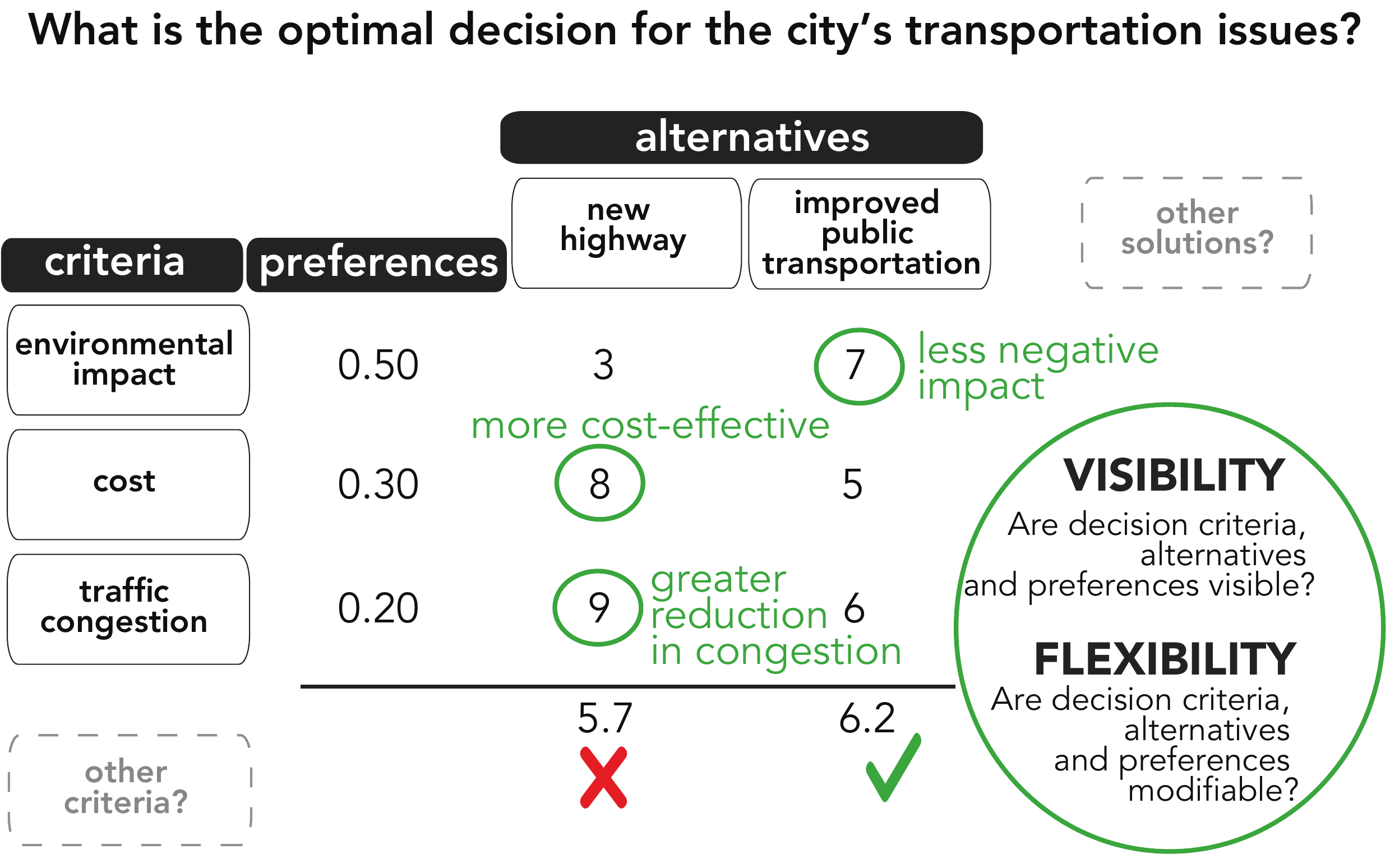}
\vspace{-0.5em}
 \caption{An illustration of a decision matrix  of a  decision maker in urban development consisting of \emph{criteria}, \emph{alternatives}, and \emph{preferences}. 
 While the new highway may be more cost-effective and reduce congestion to a greater extent, the improved public transportation has less negative environmental impact, reflecting the decision maker's preference (i.e. weight function) for environmental considerations.
} 
\vspace{-1em}
\label{fig:decisionmatrix}
\end{figure}

\subsection{Decision VISIBILITY}
\label{sec:metrics_tags_visibility}

Visualization research has studied decision making in critical application domains such as engineering design \cite{Cibulski2020PAVED} and civics \cite{Jasim2021CommunityPulse} as well as in domain-agnostic setups of personal \cite{Dy2021Improving, Dimara2018DecisionSupport}, and group  \cite{Hindalong2022GroupDecisions, Mahyar2017ConsesnsUs} decision making. A  consistent observation among these works is the need for transparency by means of direct retrieval of the decision \emph{criteria}, \emph{alternatives}, and often \emph{preferences} (or weights) as illustrated in Fig.\,\ref{fig:decisionmatrix}.

The visibility of a system's functionalities is crucial for effective interface design, as emphasized by HCI and Design expert
\cite{norman2013design, nielsen2005ten}. 
Nielsen's first usability heuristic suggests that system design should  keep users informed and provide timely feedback 
\cite{nielsen2005ten}, while Norman highlights visibility as a basic design principle.  For decision making, 
visible criteria and transparent factors enhance informed decisions, trust, and satisfaction
\cite{wang2022transparency, Handler2022ClioQuery,Jasim2022Supporting}. 
Researchers note that the lack of transparency in machine learning algorithms, obscuring decision criteria, negatively impacts users \cite{Chatzimparmpas2020Trust,Sperrle2021MLEvaluations}. Transparent computational approaches further help decision makers to figure out conflicts and differences in opinions and build common ground \cite{Mahyar2017ConsesnsUs}. 

Applying transparency to complex problems and system design is not a trivial task. One great counterexample is using \revision{reduction techniques} such as dimensionality reduction.
Dimensionality reduction  (\tag{dr}) techniques help with visual exploration of high-dimensional data by projecting the data into low-dimensional space, mostly 2D or 3D \cite{Tian2021Projection}.  For example, \cite{Talukder2020Paletteviz} implemented a  method to present the points on low-dimensional space to  represent the trade-offs of the decision space. However, despite the  important benefits of such \revision{reduction techniques}, \highlightme{decision makers see them unfavorably, even in domains with complex Pareto optimization problems because they hinder the direct
retrieval of any criteria value from any alternative \cite{Cibulski2020PAVED}.}

The importance of supporting transparency amplifies in sociopolitical decision making such as elections, urban planning, and environmental and health policies where various stakeholders have high stakes in decisions \cite{mahyar2020designing}. 
 Especially due to the inherently political nature of civic decisions, transparency in decision making becomes increasingly important \cite{Baumer2022Political}. Yet, given the size and complexity of data and supporting the visibility of features, various functionalities and criteria in complex systems remain a challenge that has not been explored yet. 

This paper examines the extent to which decision-focused visualization tools (Sec. \ref{sec:metrics_tags_decisionfocusedvis}) support transparency, particularly in terms of decision \emph{criteria}, \emph{alternatives}, and \emph{preferences}, by providing methods for making them visible and comprehensible to all stakeholders.

\begin{figure}[H]
\vspace{-1em}
 \centering 
 \includegraphics[width=0.95\columnwidth]{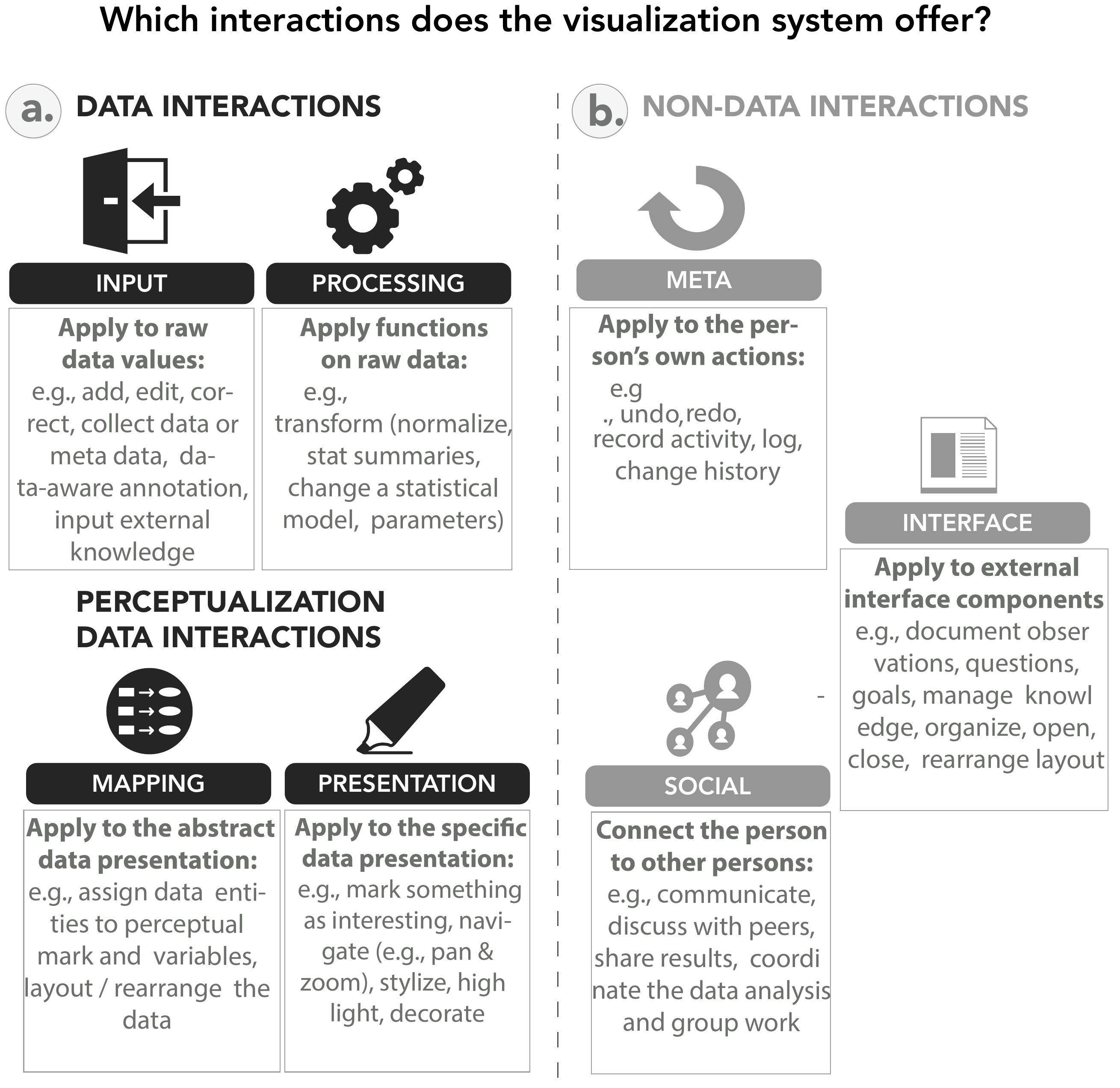}
\vspace{-1em}
 \caption{Classes of interactions of our \tag{flexibility} metric based on the  Dimara \& Perin interaction taxonomy  \cite{Dimara2020Interaction}. 
} 
\label{fig:interaction_taxonomy}
\vspace{-1em}
\end{figure}

\subsection{Decision FLEXIBILITY}
\label{sec:metrics_tags_flexibility}

Enhancing visualization interactivity has been suggested to counteract decision biases. By examining the interaction logs of decision makers, 
researchers can identify behavioral signs of cognitive bias in visual analytics \cite{wall2017warning}.
A strategy influenced by the \tag{mcdm}  method \quotes{elimination by aspects} has demonstrated its effectiveness in reducing decision biases, as it prompts the exclusion of irrelevant information from the display \cite{Dimara2019Mitigating}. Decision makers across various roles, organizational sizes, and sectors (e.g., commercial, nonprofit, \noindent health, government, education) have expressed concerns over the limited interactivity of current visualization tools, calling for more data input options and a more adaptable organization of their data within customized layouts \cite{Dimara2021Organizations}.  
We now unpack what this reported lack of flexibility entails.

\begin{figure*}
    \centering
    \includegraphics[width=\linewidth]{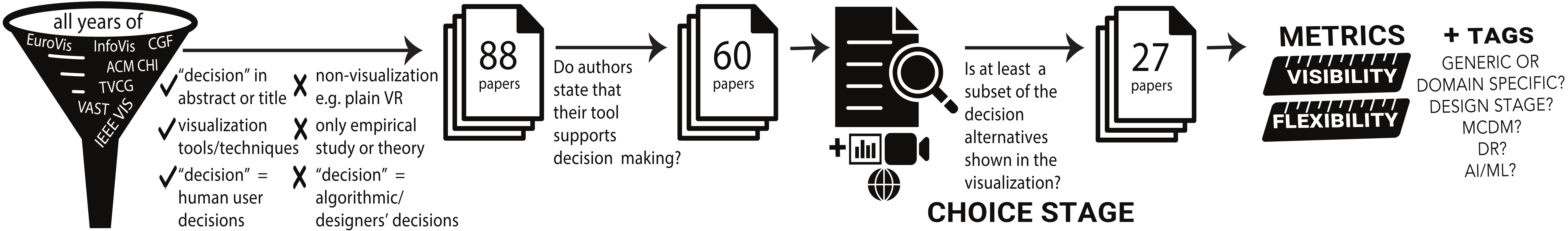}
    \vspace{-1.5em}
    \caption{Our methodology for paper collection and content analysis of \quotes{decision-focused} visualization tools in the final 27 papers. }
    \vspace{-1.5em}
    \label{fig:papercollection}
\end{figure*}

Flexibility in visualization has been connected to the degree of interactivity offered by a visualization system. Flexibility is defined as \quotes{the number of distinct, allowable actions of a person on the interface, as well as the number of interaction means with which the person can perform each action}, a concept based on Gibson's idea of affordance, which refers to \quotes{what the environment offers the individual}\,\cite{Dimara2020Interaction}. To measure the flexibility of a visualization system, researchers have proposed an interaction taxonomy that includes seven classes of interactions: \tag{input} (interactions with raw data)  or their functions), \tag{processing} (interactions with functions over raw data),
\tag{mapping} (changing or choosing data representation), 
\tag{presentation} (modifying the specific data representation), as well as  \tag{meta}, \textsc{social} and \tag{interface} (non-data interactions) \cite{Dimara2020Interaction} (see Fig.\,\ref{fig:interaction_taxonomy}).

Among these, \tag{presentation} is the most prevalent  interaction class, enabling users to alter the specific data presentation through features such as highlighting, filtering and zooming (i.e., standard \quotes{interaction techniques}  \cite{Yi2007Interaction})). 
Another class is \tag{processing}, which allows users to interact with functions over the raw data, like modifying the underlying models or  their parameters. 
\tag{mapping}, a less common class, deals with whether users can alter or even select the way data are represented. \tag{mapping} interactions, as demonstrated by sketch-based data visualization \cite{xia2018dataink}, are essential for decision makers to enable representations that align with their mental models of the decision space \cite{Dimara2021Organizations}. \tag{input} interactions, which are usually the least implemented in visualization systems  \cite{jansen2013interaction}, are particularly crucial for decision makers \cite{Dimara2021Organizations}. \tag{input} interactions involve systems permitting decision makers to incorporate their knowledge by editing or adding new criteria not present in the dataset, following design recommendations by Huron \& Willet on using visualizations as input methods \cite{Huron2021DataInput}.

\highlightme{This paper assesses visualization tools on their flexibility for decision making, encompassing all the aforementioned interaction classes. Fig.\,\ref{fig:interaction_taxonomy}.b  introduces three additional \quotes{non-data} interaction classes, with social \cite{Dimara2021Organizations} being particularly critical for group-level decisions \cite{bajracharya2018interactive}. 
Our focus is on \emph{data} comprising criteria, alternatives, and preferences, which are critical for implementing the \tag{choice} stage. The three extra interaction classes do not explicitly operate on these data elements but rather on external components \cite{Dimara2020Interaction}. Thus, we have excluded these classes from our \tag{flexibility} metric, concentrating our analysis on the interactions that directly impact the decision making process.}

To address decision makers' requests for increased interactivity \cite{Dimara2021Organizations}, we examine the extent to which current decision-focused visualization tools are designed to support flexibility by providing methods for making them accessible to all stakeholders.

\section{Methodology}
\label{sec:methodology}

In this study, we conducted a detailed qualitative content analysis of visualization tools that support decision making. 
This section outlines our paper collection process (Fig. \ref{fig:papercollection}),  coding  process, and the final derived metrics and tags.

\subsection{Paper Collection Process}
\label{sec:methodology_papercollection}

We collected papers from all years in all major visualization venues, TVCG, VIS, InfoVis, VAST, EuroVis, CGF, and CHI,
using their respective  libraries,  IEEE 
 (computer.org/csdl/), 
ACM 
(dl.acm.org),
Eurographics 
(diglib.eg.org)
and Google Scholar. 
We considered only full papers
with the keyword \quotes{decision} in the title or abstract,  as it encompasses a wide range of decision-related concepts, such as \quotes{decision making support},  \quotes{decision aid/assistance},  \quotes{multi-criteria decision making} and \quotes{decision analysis}.  This broad query aimed to capture the largest possible number of papers focusing on tools designed to support decision making.
\revision{We use the term \quotes{tool} to refer to all types of contributions of visualization solutions based on the IEEE VIS area model, such as new visual representations and techniques, systems as well as application-specific  solutions.} 
We then removed the papers where the word \quotes{decision} referred to designer decisions, algorithmic decisions, or did not contain data visualizations (most common in CHI entries). Our choice of visualization venues, search engines, and keywords are identical to a survey on visualization and decision making tools \cite{Dimara2021DecisionTasks}. However, we excluded empirical, theoretical or BELIV workshop papers  because our focus was on papers that proposed a tool.

From the initial \trackchanges{88} papers, \highlightme{we carefully reviewed each manuscript, focusing particularly on the abstract, introduction, design rationale, and conclusion. This allowed us to narrow down our selection to \trackchanges{60} papers where the authors
claimed that their tool supports decision making. Ultimately, we identified \trackchanges{27} papers on the design of \quotes{decision-focused} visualization tools for further analysis (see Fig. \ref{fig:papercollection} and Sec.\,\ref{sec:metrics_tags_decisionfocusedvis} where the term \quotes{decision-focused} was clarified).}

\subsection{Coding Process}
\label{sec:methodology_coding}

Our coding approach was primarily deductive incorporating  elements of inductive methods through iterative refinement of metric definitions and tag applications \cite{elo2008qualitative}. 
We employed a team-based coding approach \cite{macqueen1998codebook} involving five coders \highlightme{(the authors of this paper)} with distinct roles and responsibilities. The assessment of visualization tools was based on the design section of each paper, additional figures, video demos, supplementary materials (if available), and online searches for web-based versions of the tools or video demonstrations provided by the authors or found on YouTube and Vimeo.

Two independent primary coders\cite{macqueen1998codebook} were responsible for initially coding the tools using the predefined metrics of \tag{visibility} and \tag{flexibility}. 
To calculate inter-rater reliability, we determined Cohen’s Kappa \cite{cohen} after each iteration.
 While there are no strict guidelines on interpreting Cohen’s Kappa K value, a score of  K $\leq$ 0.7  is considered satisfactory for inter-coder reliability, and a score of K $\geq$ 0.75  is considered excellent \cite{statistical}. The first iteration consisted of verbal discussion and rough guidelines for scoring tools. The two coders independently scored 10 tools, obtaining a score of K = 0.55. This score represented moderate agreement, which was deemed unsatisfactory. In the next iteration, after evaluating discrepancies in scoring, the guidelines were strengthened and revised. The two coders independently scored 10 different tools, reaching a reliability score of K = 0.81, justifying the use of a single coder to review the remaining tools.
 A different lead coder \cite{macqueen1998codebook} reviewed all the tools coded by the independent coders, expanded the metric applications, and applied the additional tags. This lead coder was not independent, as they had seen the scores of the independent coders. Two consulting coders reviewed tool subsets and met regularly with the other coders \cite{macqueen1998codebook}  to discuss unclear points, refinements of tags and metrics, and provide insights.

\subsection{Final Tags \& Metrics Scoring System}
\label{sec:methodology_metrics_tags}

This section presents the final specifications for tags and metrics based on our 
analysis of the  literature from decision sciences and visualization (Sec. \ref{sec:metrics_tags}) and our content analysis and coding process (Sec. \ref{sec:methodology_coding}). 
Our analysis uses \trackchanges{23} tags. The first tag is the \tag{dm-goal} (\{Yes, No\}), determining if the authors claim their tool supports decision making (applied to \trackchanges{88} tools). If \tag{dm-goal} is 'Yes,' the decision making stage tags (See definitions in Sec. \ref{sec:metrics_tags}) are applied to the remaining
\trackchanges{60} tools: \tag{intelligence}, \tag{design}, and \tag{choice} (\{Yes, No\} for each).  If \tag{choice} is 'Yes,' the visualization design assessment tags are applied to the \trackchanges{27} tools. These tags are subdivided into three categories:
\begin{enumerate}
    \item Visibility Tags: \emph{criteria}, \emph{alternatives}, and \emph{preferences} (\{Yes, Partly, No\} for each), assessing the visibility of these components in the visualization.
    \item Flexibility Tags: \tag{input}, \tag{mapping}, \tag{processing}, and \tag{presentation} (\{Yes, No\} for each of the \emph{criteria}, \emph{alternatives}, and \emph{preferences}), evaluating the tool's modifiability of these elements.
    See interaction classes definitions in Sec.\,\ \ref{sec:metrics_tags_flexibility}. 
    \item Other Tags: \tag{mcdm} (\{Yes, No\}), indicating if the tool supports multi-criteria decision making; \tag{dr/da} (\{Yes, No\}), determining if the tool offers \revision{dimensionality reduction or criteria aggregation;} \tag{ai} (\{Yes, No\}), denoting if the choice stage is aided by AI or ML; and \tag{generalizability} (\{Yes, No\}), identifying if the authors claim that their tool applies to generic datasets.
\end{enumerate}

\revision{To identify the user allowable tasks, the coders used the definitions from the interaction framework reviewed  in Sec.\,\ref{sec:metrics_tags_flexibility}, but  several other interaction taxonomies could have been used (see \cite{Dimara2020Interaction} for a review).}
These tasks include \tag{input}, where the decision maker can add, edit, or delete alternatives (self-decided or based on a recommendation), criteria, and preferences (exact numbers or weight approximations). The decision maker can also perform \tag{mapping} to switch to or create a new representation of alternatives, criteria, or preferences. In \tag{processing}, the decision maker can modify or add new functions for alternatives (e.g., a new alternatives creation model), criteria (e.g., aggregation or model parameters), and preferences (e.g., \tag{mcdm} weight function). Finally, the decision maker can adjust the \tag{presentation} of alternatives, criteria, and preferences through specific representations (e.g., select, highlight, navigate, or zoom).

\highlightme{
Our simplified coding scheme with three  \tag{visibility} levels, results from the detail level in the tool descriptions from some papers. Coders couldn't extract more granularity due to insufficient explanations. Future research could benefit from additional granularity in metrics, especially if used during the design phase of a new visualization.}

Next, we discuss the scoring system of our metrics derived from the tags \tag{visibility} and \tag{flexibility}. The \tag{visibility} metric ranges from \trackchanges{0 to 12} and assigns a score of \trackchanges{4} for 'Yes,' \trackchanges{2} for 'Partly,' and \trackchanges{0} for 'No' for each of the three visibility tags (e.g., for \emph{criteria}). \highlightme{The sum of these scores for all visibility tags yields the total \tag{visibility} score.} The \tag{flexibility} metric also ranges from \trackchanges{0 to 12}, with each 'Yes' for a specific class (e.g., \tag{input}) and decision element (e.g., \emph{criteria}) receiving a score of \trackchanges{1}. \highlightme{The sum of these scores for all flexibility tags yields the total \tag{flexibility} score.}

\highlightme{We note that the numerical scores assigned to \tag{flexibility} and \tag{visibility} metrics do not carry any inherent meaning, and their absolute values should not be used for direct comparison.
To prevent any misinterpretation of the importance of one metric over the other, we assigned a maximum of \trackchanges{12} points to both \tag{flexibility} and \tag{visibility}.  These scores provide a means to compare the relative performance of tools within their respective categories. For instance, a higher \tag{flexibility}  score may suggest a tool supports a larger number of interaction classes or more decision elements. Furthermore, tools with identical scores may still differ in the specific areas of \tag{flexibility}  or \tag{visibility} in which they excel. For example, one tool may offer more \tag{flexibility} in criteria, while another may have greater \tag{flexibility} in weights.}

\begin{figure*}
 \vspace{-1.5em}
 \centering 

\includegraphics[width=\linewidth]{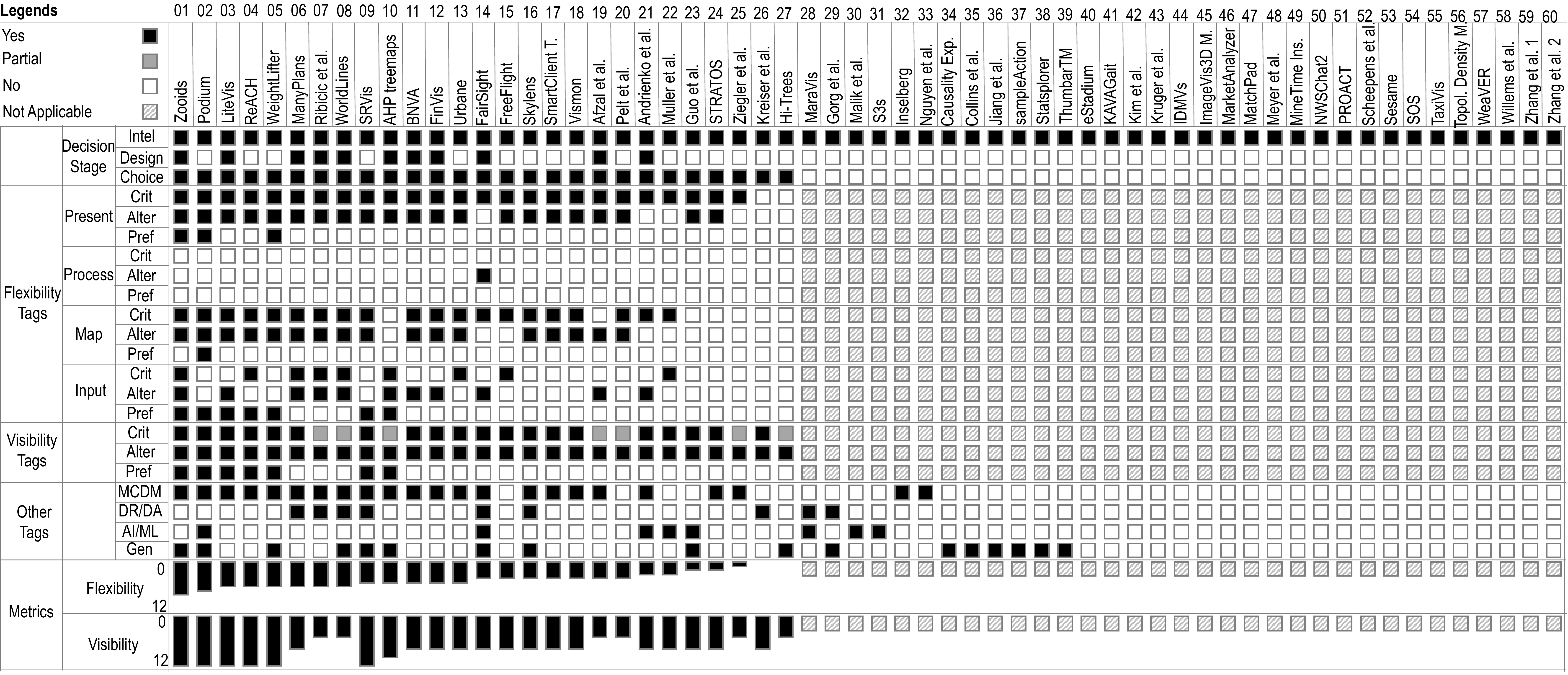}
 \vspace{-1.5em}
 \caption{
\revision{Summary of the 60 visualization tools explicitly designed to facilitate decision making, categorized based on their support for each decision making stage, flexibility, visibility, and additional tags and metrics. Notably, only 27 tools support the \tag{choice}  and 11 the \tag{design} stage.}
} 
\vspace{-2em}
\label{fig:dmtable}
\end{figure*}

\begin{figure*}
  \centering
  \includegraphics[width=0.95\linewidth]{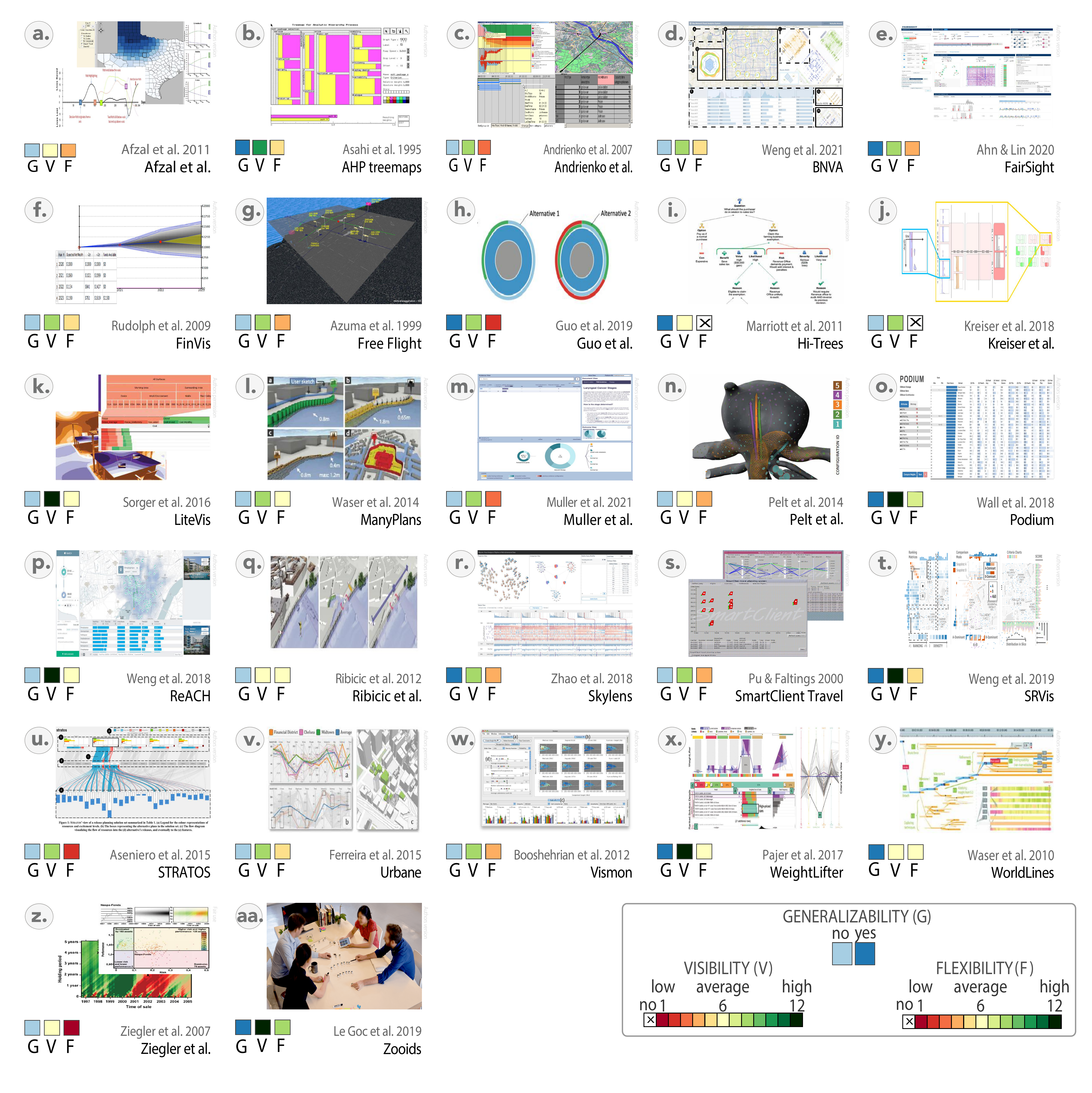}
\caption{ 
A mosaic of representative figures for the \trackchanges{27} decision-focused tools supporting the \tag{choice} stage, arranged alphabetically. The diverging color scale highlights the \tag{visibility} and \tag{flexibility} scores, with greener shades indicating above-average scores (i.e., 6/12) and red shades indicating below-average scores. In Sec.\,\ \ref{sec:metrics_tags}, we discuss how our metrics (\tag{visibility} and \tag{flexibility}) and our analysis per decision making stages are grounded in decision sciences literature.  The presence of a dark blue G square denotes \tag{generalizability}, 
indicating that the authors suggest their tool can be applied to generic datasets, while a light blue tag suggests the tool is more specific in its application. This mosaic aims to provide a comprehensive visual overview of how visualization design supports the choice stage in decision making processes.
However, it is important to note that these scores simply indicate the number of various features offered by each tool, and should not be interpreted as negative or positive judgments. In certain domain applications, having too many features may hinder efficiency and understandability.
} 
\label{fig:dmtools_mosaic}
\end{figure*}

\section{Results}
\label{sec:results}

This section presents the results of our content analysis of the \trackchanges{60} visualization tools designed to support decision making,  from the \trackchanges{88} decision-relevant papers published in major visualization journals and conferences over all available years. Fig.\,\ref{fig:dmtable} displays our tagging, as described in Sec.\,\ref{sec:methodology} (black square for 'Yes', white for 'No') and metric scores (black bar charts). Fig.\,\ref{fig:dmtools_mosaic} showcases representative figures of the \trackchanges{27} decision-focused tools that support the \tag{choice} stage, using a diverging color scale. 
The complete curated list of the \trackchanges{88} tools is available in the supplementary materials, for which we have provided the OSF link in the abstract section. 
For each visualization tool, we will use the notation ([citation], 
\revision{column number
in Fig.\,\ref{fig:dmtable}}, letter in Fig,\,\ref{fig:dmtools_mosaic}).

\subsection{How visible are the decision-focused visualizations?}
\label{sec:results_visibility}

\subsubsection{Visibility of Decision Alternatives}
\label{sec:results_visibility_alternatives}

All \trackchanges{27} decision-focused tools show alternatives in a fully visible manner. While most of them facilitate the comparison of alternatives on the same screen,  
Andrienko \emph{et al.} 
(\cite{Andrienko2007Evacuation}, \trackchanges{21}, \trackchanges{c})
is an exception, which presents alternatives sequentially. Andrienko \emph{et al.} represents the parameters of an evacuation plan, and a planner can explore each criterion and monitor how the new plans perform based on the number of people rescued, the capacity of the vehicles at any time, and the status of nearby cities in which people are evacuated. The sequential presentation allows for a step-by-step exploration of each dimension and the creation of a new plan based on the changing parameters. Overall, the high level of visibility of alternatives in the reviewed tools indicates that visualization designers recognize the importance of making all decision alternatives clearly visible to decision makers.

Besides visibly presenting alternatives, some tools provide features to enhance scalability and user experience with numerous alternatives.
For example, 
BNVA
(\cite{Weng2020Bus}, \trackchanges{11}, \trackchanges{d}) 
allows users to expand and collapse alternative sets based on shared features, such as bus routes with shared stops, to improve the scalability of the system for larger datasets. Similarly, 
ManyPlans
(\cite{Waser2014ManyPlans}, \trackchanges{6}, \trackchanges{l})
and 
Ribicic \emph{et al.}
(\cite{Ribicic2012Sketching}, \trackchanges{7}, \trackchanges{q})
enable users to collapse different connected branches into a single track while still providing the ability to expand the original branches, enhancing the user experience for growing alternative sets.

\subsubsection{Visibility of Decision Criteria}
\label{sec:results_visibility_criteria}

Out of the \trackchanges{27} decision making tools we analyzed, criteria are fully visible in \trackchanges{20}, while the remaining tools only partially display criteria due to deliberate design choices. 
Afzal et al. 
(\cite{Afzal2011Epidemic}, \trackchanges{19}, \trackchanges{a})
acknowledged their design limitation, as users can monitor only one criterion on the decision history tree's y-axis. 
Ribicic et al.
(\cite{Ribicic2012Sketching}, \trackchanges{7}, \trackchanges{q})
and 
WorldLines 
(\cite{Waser2010WorldLines}, \trackchanges{8}, \trackchanges{y})
showed a subset of criteria in their evaluation view, even though the flood manager user requested to balance cost, time, and effectiveness.
AHP Treemaps 
(\cite{Asahi1995Treemaps}, \trackchanges{10}, \trackchanges{b})
allows users to see only the title of criteria in a treemap and not their values, while 
Hi-Trees
(\cite{Marriott2011HiTrees}, \trackchanges{27}, \trackchanges{i})
displays facts about alternatives (cons, pros, etc.) 
without a clear set of criteria. 
Pelt et al. 
(\cite{Pelt2014Blood}, \trackchanges{20}, \trackchanges{n})
show only treatment-focused criteria on their 3D layered model view, and 
Ziegler et al.
(\cite{Ziegler2007Relevance}, \trackchanges{25}, \trackchanges{z})
present risk and  assets performance on a Dominance Plot.

Sec.\,\ref{sec:metrics_tags_visibility} suggested that criteria obscuring concerns are often associated with \revision{dimensionality reduction or data aggregation}. However, some reviewed tools with the \tag{dr/da} tag, 
like 
SRVis 
(\cite{Weng2019SRVis}, \trackchanges{9}, \trackchanges{t})
and 
Skylens 
(\cite{Zhao2018SkyLens}, \trackchanges{16}, \trackchanges{r}),
facilitate the exploration of similarities between alternatives while still maintaining full visibility of criteria. In addition to the projection view that shows alternatives in 2D space, they offer other representations such as tabular views (SRVis, Skylens) and radar charts (Skylens) displaying all criteria transparently.
Moreover, there were tools with data aggregation that offered full criteria visibility, like 
FairSight 
(\cite{Ahn2019FairSight}, \trackchanges{14}, \trackchanges{e})
and 
Kreiser \emph{et al.},
(\cite{Kreiser2018DecisionGraph}, \trackchanges{26}, \trackchanges{j})
.
Although FairSight only shows aggregated utility and fairness measures of each ranking on the ranking list view, a user can see the details of all criteria in the feature inspection view. Similarly, although Kreiser \emph{et al.} acknowledged the importance of focusing user attention on treatment-specific information, their design choice kept both decision-critical aggregated swallow data and individual  data on the same screen while highlighting decision-critical information with highly saturated colored regions.

\subsubsection{Visibility of Decision Preferences}
\label{sec:results_visibility_preferences}

Despite most tools displaying alternatives and criteria with full visibility, 
as discussed in Sec.\,\ref{sec:results_visibility_alternatives} and Sec.\,\ref{sec:results_visibility_criteria}, 
only \trackchanges{7} of them fully show decision preferences. 
The most common method, found in \trackchanges{5} of these \trackchanges{7} tools,
including 
Podium
(\cite{Wall2018Podium}, \trackchanges{2}, \trackchanges{o})
, 
LiteVis
(\cite{Sorger2016Litevis}, \trackchanges{3}, \trackchanges{k})
, 
ReACH 
(\cite{Weng2018Homefinder}, \trackchanges{4}, \trackchanges{p})
,
WeightLifter
(\cite{Pajer2017Weightlifter}, \trackchanges{5}, \trackchanges{x})
and 
SRVis 
(\cite{Weng2019SRVis}, \trackchanges{9}, \trackchanges{t})
,
enables users to rank alternatives and observe the outcome weights on stacked bar graphs.  Notably, WeightLifter  also features a unique triangular visual representation that facilitates sensitivity analysis of weight changes. 
In contrast, 
AHP treemaps 
(\cite{Asahi1995Treemaps}, \trackchanges{10}, \trackchanges{b})
and 
Zooids 
(\cite{LeGoc2019Dynamic}, \trackchanges{1}, \trackchanges{aa})
present decision preferences using different representations. 
AHP treemaps allows users to input their decision problem, criteria titles, and alternatives in a hierarchical treemap structure with entities coded as rectangles. Decision makers can adjust rectangle sizes to express their preference for different alternatives and criteria. 
Zooids, which are mini robots, demonstrated a use case for an admission committee selecting the best candidate to enroll. Committee members can load candidate data into candidate zooids by placing an empty zooid onto candidate data on a tablet screen. Similarly, they can place a magnet zooid on a criterion on the tablet. To assign different weights to various criteria, users adjust the attraction force of magnet zooids by rotating them clockwise or counterclockwise, encoding different criteria values and eliciting preference weights for each criterion.
\subsection{How flexible are the decision-focused visualizations?}

\label{sec:results_flexibility}
 
\subsubsection{Flexibility of Decision Alternatives}
\label{sec:results_flexibility_alternatives}

Concerning the \tag{presentation} of alternatives, our analysis shows that all tools, except \trackchanges{6},
enable users to modify the alternatives' presentation using at least one interaction medium (like hovering, brushing and linking, selecting and highlighting, filtering, adjusting visual marks, or panning and zooming).
Although our binary scoring solely captures the availability of at least one of these interaction techniques, our further design analysis reveals that some tools provide multiple techniques 
(e.g., 
WeightLifter
(\cite{Pajer2017Weightlifter}, \trackchanges{5}, \trackchanges{x})
,
WorldLines
(\cite{Waser2010WorldLines}, \trackchanges{8}, \trackchanges{y})
,
Pelt et al.
(\cite{Pelt2014Blood}, \trackchanges{20}, \trackchanges{n})
.
Tools lacking alternatives presentation flexibility, like
FairSight
(\cite{Ahn2019FairSight}, \trackchanges{14}, \trackchanges{e})
, show alternatives rankings in a tabular view with static sliders representing different criteria utility and fairness values, without any interaction for manipulating the presentation of rankings.

FairSight
(\cite{Ahn2019FairSight}, \trackchanges{14}, \trackchanges{e})
is the only tool enabling interaction at the \tag{processing} level, allowing to modify functions or add models to create alternatives (see Sec.,\ref{sec:methodology_metrics_tags}).
While \trackchanges{11} tools
offer ways to create new alternatives using existing functions (e.g.,
FinVis
(\cite{Rudolph2009Finvis}, \trackchanges{12}, \trackchanges{f})
uses a menu to input criteria values and a button to add the new alternative to the tabular view and line graph),
they do not provide an interactive way to modify or add another function.

Like the presentation-level flexibility for alternatives, we found that most tools provide users the ability to change \tag{mapping} by switching or creating new visual representations of 
alternatives. 
For example, 
Pelt et al.
(\cite{Pelt2014Blood}, \trackchanges{20}, \trackchanges{n})
built a visualization system in which a zoomable multilayered 3D model of a brain vessel shows several stent configurations (alternatives) in different visual representations, such as plain disks, flower glyphs, or radar charts at each zoom level.
FinVis
(\cite{Rudolph2009Finvis}, \trackchanges{12}, \trackchanges{f})
enables users to switch between a tabular view and a line chart to monitor alternative investments in different visual representations. 
In contrast to Pelt et al., most tools assist flexibility of alternatives mapping only by switching between different windows 
(e.g., 
SmartClient Travels
(\cite{Pu2000SmartClient}, \trackchanges{17}, \trackchanges{s})
,
ReACH 
(\cite{Weng2018Homefinder}, \trackchanges{4}, \trackchanges{p})
,
Podium
(\cite{Wall2018Podium}, \trackchanges{2}, \trackchanges{o})
,
ManyPlans
(\cite{Waser2014ManyPlans}, \trackchanges{6}, \trackchanges{l})
). 
Furthermore, 
Zooids 
(\cite{LeGoc2019Dynamic}, \trackchanges{1}, \trackchanges{aa})
assists users with a broad range of possibilities, allowing to place mini robots in any configuration
, such as freely placing them on the table and using magnet zooids to see the position of alternative zooids with respect to magnet (i.e., criteria) or creating a scatterplot formation of alternatives.

On alternatives \tag{input}, we found only \trackchanges{11} tools allowing to interactively add alternatives, crucial for the \tag{design} decision stage. Among them,
LiteVis
(\cite{Sorger2016Litevis}, \trackchanges{3}, \trackchanges{k})
,
ManyPlans
(\cite{Waser2014ManyPlans}, \trackchanges{6}, \trackchanges{l})
,
Ribicic et al.
(\cite{Ribicic2012Sketching}, \trackchanges{7}, \trackchanges{q})
and
WorldLines
(\cite{Waser2010WorldLines}, \trackchanges{8}, \trackchanges{y})
support alternative creation by clicking nodes on a tree view, inputting parameters using 2D or 3D sketch views, or selecting a region in a 3D simulation view and inputting simulation parameters in another steering view.
FairSight
(\cite{Ahn2019FairSight}, \trackchanges{14}, \trackchanges{e})
uses a machine-learning algorithm where users select different parameter priorities to create ranking alternatives from a candidate list, enabling fair and unbiased decisions.
Afzal et al.
(\cite{Afzal2011Epidemic}, \trackchanges{19}, \trackchanges{a})
allow inputting various epidemic countermeasures using a pop-up menu to enter values for individual measures via text fields. 
Zooids
(\cite{LeGoc2019Dynamic}, \trackchanges{1}, \trackchanges{aa})
offers a highly flexible method for creating alternatives by placing a zooid on a tablet's touch screen to load a candidate's data.

\subsubsection{Flexibility of Decision Criteria}
\label{sec:results_flexibility_criteria}

Almost all tools offer flexible criteria at the \tag{presentation} level, with the exceptions of
Kreiser et al.
(\cite{Kreiser2018DecisionGraph}, \trackchanges{26}, \trackchanges{j})
and
Hi-Trees
(\cite{Marriott2011HiTrees}, \trackchanges{27}, \trackchanges{i})
.
The remaining tools enable criteria flexibility at the presentation level by simply highlighting STRATOS (\cite{Aseniero2015Stratos}, 24, u), navigating
(e.g., changing the angle of the 3D model in
Pelt et al.
(\cite{Pelt2014Blood}, \trackchanges{20}, \trackchanges{n})
),
filtering criteria values using sliders
(e.g.,
Vismon
(\cite{Booshehrian2012Vismon}, \trackchanges{18}, \trackchanges{w}); Free Flight (\cite{azuma1999}, 15, g)),
Guo et al
(\cite{Guo2019Event}, \trackchanges{23}, \trackchanges{i})
),
checkboxes in
Afzal et al.
(\cite{Afzal2011Epidemic}, \trackchanges{19}, \trackchanges{a}),
slider bar on a histogram in
BNVA
(\cite{Weng2020Bus}, \trackchanges{11}, \trackchanges{d}),
dragging and dropping criteria values to change their order
in
AHP Treemaps
(\cite{Asahi1995Treemaps}, \trackchanges{10}, \trackchanges{b}),
and brushing on parallel coordinates
(e.g.,
WeightLifter
(\cite{Pajer2017Weightlifter}, \trackchanges{5}, \trackchanges{x}));
adding stickers onto physical mini robots
in
Zooids
(\cite{LeGoc2019Dynamic}, \trackchanges{1}, \trackchanges{aa}),
collapsing and expanding line ensembles for different simulation parameters in
ManyPlans
(\cite{Waser2014ManyPlans}, \trackchanges{6}, \trackchanges{l})
and
Ribicic \emph{et al.}
(\cite{Ribicic2012Sketching}, \trackchanges{7}, \trackchanges{q}).
In contrast to alternatives, where
FairSight
(\cite{Ahn2019FairSight}, \trackchanges{14}, \trackchanges{e})
offers flexible alternatives at the processing level, we did not find any tools that assist \tag{processing} level interaction of criteria.

On \tag{mapping}, most tools are interactive via switching to different windows (e.g.,
Podium
(\cite{Wall2018Podium}, \trackchanges{2}, \trackchanges{o}),
LiteVis
(\cite{Sorger2016Litevis}, \trackchanges{3}, \trackchanges{k}),
ReACH
(\cite{Weng2018Homefinder}, \trackchanges{4}, \trackchanges{p})
,
BNVA
(\cite{Weng2020Bus}, \trackchanges{11}, \trackchanges{d})).
For some tools, it was possible to switch to other visual representations of criteria by using a zoom function
(%
Pelt et al.
(\cite{Pelt2014Blood}, \trackchanges{20}, \trackchanges{n})).
Zooids
(\cite{LeGoc2019Dynamic}, \trackchanges{1}, \trackchanges{aa})
enables users to pick up and place robots anywhere in any form of visual representation, demonstrating the flexibility of physicalization to facilitate decision makers in evaluating criteria from different perspectives.

Only \trackchanges{9} tools support flexible criteria \tag{input}.
ReACH
(\cite{Weng2018Homefinder}, \trackchanges{4}, \trackchanges{p})
users can input reachability constraints, such as departure and arrival times, affecting reachability scores visualized as a heatmap.
ManyPlans
(\cite{Waser2014ManyPlans}, \trackchanges{6}, \trackchanges{l})
,
Ribicic \emph{et al.}
(\cite{Ribicic2012Sketching}, \trackchanges{7}, \trackchanges{q})
and
WorldLines
(\cite{Waser2010WorldLines}, \trackchanges{8}, \trackchanges{y})
users can modify criteria for all alternatives, like  water velocity or embankment type.
Urbane
(\cite{Ferreira2015Urbane}, \trackchanges{13}, \trackchanges{v})
users can adjust criteria, such as building location and height, using select and steer options in 3D mode to evaluate urban development suitability in each region.

\subsubsection{Flexibility of Decision Preferences}
\label{sec:results_flexibility_preferences}

We found limited flexibility of preferences compared to alternatives and criteria. At the \tag{presentation} level, only \trackchanges{3} tools enable flexibility of preferences.
Zooids
(\cite{LeGoc2019Dynamic}, \trackchanges{1}, \trackchanges{aa})
lets users modify the attraction force of a magnet zooid (criteria) and observe changes in the light's brightness.
Podium
(\cite{Wall2018Podium}, \trackchanges{2}, \trackchanges{o}),
allows toggling the visibility of bars representing criterion weight.
In WeightLifter
(\cite{Pajer2017Weightlifter}, \trackchanges{5}, \trackchanges{x})
, users can filter weight values (e.g., setting a minimum weight for a particular criterion). No tools were found that support \tag{processing}-level interaction for preferences. Importantly, only Podium provides \tag{mapping} flexibility, enabling users to view preferences in various visual representations by switching between views (e.g., vertical lines in tabular view and red/green bars in the control panel under the attributes tag).

Only \trackchanges{7} tools support \tag{input} flexibility of preferences.
Most have a ranking view and feature stacked bars in a tabular layout, allowing to add preferences by changing the size of rectangles representing criteria
(e.g.,
LiteVis
(\cite{Sorger2016Litevis}, \trackchanges{3}, \trackchanges{k})
,
ReACH
(\cite{Weng2018Homefinder}, \trackchanges{4}, \trackchanges{p})
,
WeightLifter
(\cite{Pajer2017Weightlifter}, \trackchanges{5}, \trackchanges{x})
SRVis
(\cite{Weng2019SRVis}, \trackchanges{9}, \trackchanges{t}))
. Instead, to input preferences,
Podium
(\cite{Wall2018Podium}, \trackchanges{2}, \trackchanges{o})
,
allows to drag and drop rows in the tabular view, and the model calculates weights based on users' manual ranking. Users can also change their preferences by adjusting the size of horizontal bars in the control panel.
With
Zooids
(\cite{LeGoc2019Dynamic}, \trackchanges{1}, \trackchanges{aa})
, users can modify the attraction force of magnet zooids to input their preferences for a criterion.
AHP treemaps
(\cite{Asahi1995Treemaps}, \trackchanges{10}, \trackchanges{b})
enable users to input their preferences for criteria and alternatives using the hook and pump tools.

\subsection{AI Assistance,  Domains \& Lack of Choice Support}

Only \trackchanges{5} decision-focused tools provide algorithmic assistance
(e.g., for preference elicitation, ranking, scheduling, event sequence prediction, and treatment recommendation). 
Podium
(\cite{Wall2018Podium}, \trackchanges{2}, \trackchanges{o})
utilizes Ranking SVM to facilitate users' preference elicitation process. Podium users can drag and drop rows to manually rank the alternatives,
and the model calculates inferred weights for criteria.
Andrienko \emph{et al.}
(\cite{Andrienko2007Evacuation}, \trackchanges{21}, \trackchanges{c})
use a genetic algorithm to automatically create an evacuation schedule and
FairSight
(\cite{Ahn2019FairSight}, \trackchanges{14}, \trackchanges{e})
implemented their own machine learning model based on top-k ranking to enable users to identify possible biases (e.g., on gender) and also select different ranking algorithms such as RankSVM, Logistic Regression, and SVM.
Muller et al.
(\cite{muller2021}, \trackchanges{22}, \trackchanges{m})
utilize BNs to create the optimal treatment option for cancer treatment.
Guo et al
(\cite{Guo2019Event}, \trackchanges{23}, \trackchanges{i})
utilize TRNN to predict the event sequence, the outcome of which is later shown in a disk glyph.
Most of these tools lack the ability to enable users to change model parameters to support the \tag{design} stage of their decision making process.

Different domains, as shown in Fig.\,\ref{fig:dmtable} Gen,  exhibit various design patterns and focus on different aspects of the decision making process.
 In domains where trade-off analysis is critical, more emphasis is placed on the \tag{design} stage, and interactive sketching is utilized to explore alternative solutions. 
Examples include flood management (
Ribicic \emph{et al.}
(\cite{Ribicic2012Sketching}, \trackchanges{7}, \trackchanges{q})
,
ManyPlans
(\cite{Waser2014ManyPlans}, \trackchanges{6}, \trackchanges{l})
),
urban planning (
Urbane
(\cite{Ferreira2015Urbane}, \trackchanges{13}, \trackchanges{v})
), and
lighting design (
LiteVis
(\cite{Sorger2016Litevis}, \trackchanges{3}, \trackchanges{k})
). 
In the finance domain, tools such as 
FinVis
(\cite{Rudolph2009Finvis}, \trackchanges{12}, \trackchanges{f})
and
Ziegler et al.
(\cite{Ziegler2007Relevance}, \trackchanges{25}, \trackchanges{z})
, tend to have simpler designs focusing on selecting assets and portfolios. In the health domain, designers prioritized displaying decision-critical criteria. Understanding domain-specific challenges guides the creation of tailored decision making tools. Despite the importance of focusing doctors on key information, the paper didn't clarify why access to all criteria was restricted. 
Interestingly, our top-2 flexible tools are domain-general tools (Zooids and Podium).

The other \trackchanges{33} tools, which assert support for decision making, focus solely on data exploration and analysis.
For instance, MatchPad offers real-time analysis of match statistics and events on a tablet interface, claiming to aid coaches in decision making. While exploring data is useful, decision makers may require more support in the \tag{design} and \tag{choice} stages for informed decisions.
Similarly, tools like Zhang \emph{et al.} and TaxiVis support urban data analysis but lack alternatives. Nguyen \emph{et al.} approaches decision making more closely by facilitating multi-faceted decision making in cybersecurity, but ultimately, focuses on identifying and investigating unusual action sequences.

\section{Discussion}

Our extensive search for visualization tools designed to support decision making uncovered a scarcity of decision-focused tools (see Sec.~\ref{sec:metrics_tags_decisionfocusedvis}). Despite the vastness of the decision making domain,  
we were surprised to find 
a number of tools in our corpus comparable to that of visualization surveys on more specialized and recent topics, such as Bitcoin visualization~\cite{tovanich2019visualization}. 
In these tools, we found that support for decision preferences and flexibility was inadequate, particularly in terms of input and processing interactions, as well as preferences mapping, despite the visibility of alternatives and criteria. Next, we delve deeper into our findings and  their implications for decision makers.

\subsection{Implications of Missing Features in Decision Tools}
\color{black}
\noindent
\revision{Missing certain features in visualization tools can have implications for decision effectiveness. Our findings revealed limited support for automating certain parts of the decision making process}, including AI assistance, dimensionality reduction, and decision guidance. This is surprising given the complexity of modern decisions, the abundance of data, and the increasing number of criteria to consider. The integration of AI in visualization systems has been widely explored \cite{Chatzimparmpas2020Trust, Sperrle2021MLEvaluations}, highlighting its potential also for decision making. Future research can focus on facilitating the choice stage, especially when dealing with numerous alternatives or attributes, by incorporating recommendations with user control via enhanced flexibility levels  of processing interactions. This in turn will help enhance the scalability of tools. For instance, categorical colors can be used to encode decision criteria, such as stacked bar visualizations \cite{Pajer2017Weightlifter, Sorger2016Litevis}.
Additionally, AI assistance in preference elicitation, an area where MCDM research can provide valuable insights \cite{Zavadskas2011MCDM}, can shed light on  the complexity of the preference elicitation process. Assisting decision makers in this area not only benefits their decision making process but also supports the design of evaluation protocols that rely on ground truth metrics \cite{Dimara2018DecisionSupport}. Decision strategies also need to be recommended to decision makers \cite{Dimara2019Mitigating}. Interestingly, our review did not identify any tool that suggests a way of deciding, an area where MCDM could offer valuable guidance.
\revision{Inadequate automation support for human-driven decision making may fuel arguments for complete automation, given the potential for human errors due to data complexity.}

\revision{In addition to the absence of automation support, the lack of input features in many of these tools further restricts opportunities for}  decision makers to incorporate their own knowledge \cite{DataHunches2023}, despite prior user feedback demanding such features in existing works \cite{mccurdy2018framework, Dimara2018DecisionSupport}. \revision{Furthermore, the scarcity of input features also affects the stages of decision making they support,} particularly the design stage, which is currently only addressed by a few tools. \revision{During this stage, scenario simulations necessitate the input of new alternatives}, potentially explaining the limited number of tools that incorporate simulations and 3D model views showcasing their corresponding effects~\cite{Waser2010WorldLines, Waser2014ManyPlans, Ribicic2012Sketching, Ferreira2015Urbane}.
To address this limitation, visualization research could explore the use of visualization as data inputs, as suggested by Huron and Willet~\cite{Huron2021DataInput}.

\revision{Furthermore, it's not just whether certain features are supported but also their quality. For example, we observed standard mapping features} that allow decision makers to switch between existing and effective in terms of precision representations, like tabular layouts~\cite{bartram2021untidy,Dimara2018DecisionSupport}, stacked bars~\cite{siirtola2014bars,Gratzl2013Lineup}, parallel coordinates~\cite{gettinger2013comparison, Dimara2018DecisionSupport}, and spatial layouts for specific domains \cite{Weng2020Bus}.
Yet, these representations were predetermined by the designers. Decision makers could only view their decision space through the eyes of the designer. 
In previous research, when asked to draw decision-related information, decision makers seemed to create different layouts, like flow diagrams, pros and cons lists, and unique trees~\cite{Dimara2021Organizations}.
This underlines the importance of visualization research investigating how to allow decision makers to view their decision space within their own \emph{mental model}~\cite{Liu2010Mental},
e.g., via abstracted components, sketching, or physical manipulation like Zooids \cite{LeGoc2019Dynamic}. Such flexibility would be crucial for the decision making process, which, besides support from formal decision analysis methods \cite{keeney1982decision}, can be enhanced with brainstorming, collaboration, synthesis, and planning \revision{features} \cite{chermack2011scenario,klein2008naturalistic}.

\subsection{Extending Decision Support Features: An Example}
\color{black}

\noindent
\revision{
In Sec. \ref{sec:introduction} and \ref{sec:metrics_tags}, we discussed the information processing requirements of decision makers, using the city planner scenario as an example (Fig.~\ref{fig:decisionmatrix}). Now, let's illustrate how visualization tools can be extended to incorporate decision support features, continuing with the city planner scenario.
In our scenario, the decision maker, using the urban planning-relevant tool BNVA (from our review, Fig.~\ref{fig:dmtable}.11, Fig.~\ref{fig:dmtools_mosaic}.d), addresses a traffic issue.
Currently, the tool already enables a good range of features, including support for all decision stages and an interactive map for determining the impact of different bus route implementations on traffic conditions.}

\revision{
To enhance BNVA, we propose adding input functionalities to encompass a broader range of alternative solutions, such as new bike lanes or highways. This expansion would also enable city planners to include decision criteria beyond pre-defined sets, permitting iterative adjustments to solution rankings based on new evaluation criteria.
To create a flexible decision space, BNVA could transcend traditional formats like tabular layouts and embrace sketch-based capabilities \cite{xia2018dataink}. 
Mirroring the flexibility offered by Zooids \cite{LeGoc2019Dynamic}, within a digital whiteboard, 
decision makers can project their mental models onto visual representations, annotating and comparing solutions, such as constructing a new highway or improving public transportation.
Planners could then employ data-aware sticky notes to annotate the new highway solution with remarks like \quotes{reduced travel times}
or \quotes{increased road capacity}.
Infusing BNVA with automation capabilities could guide planners further by spotlighting solutions aligning with their preferences, akin to Podium \cite{Wall2018Podium},  facilitating revisions within the decision space.}

\revision{However, it is important to note that not all decision making tools must incorporate each feature identified by our review's tags and metrics to be considered effective. Understanding the specific needs and context of decision making is crucial. In some cases, simplicity and speed may be paramount, while in others,  complex features are needed to address domain complexity. However, we found papers often lack discussions of design trade-offs in the context of decision making tasks. The next section highlights the importance of the underlying design process to identify decision makers' needs in various contexts.}

\subsection{Design Process for Tailored Decision Tools}
\color{black}

\noindent\revision{Effective design involves not only determining which features to include but also following a suitable methodology to derive and refine those features. For that, designers need to be able to concretize the decision problem at hand. However,} during our survey, we encountered difficulties in extracting concrete information about the decision elements, such as alternatives, criteria, and preferences from most papers. We are concerned that this may reflect a less tailored design process followed for the creation of decision making tools.
\revision{Tool designs should be influenced by user profiles, encompassing decision makers with diverse goals, domains, and varying data analysis and visualization proficiency \cite{Baumer2022Political}. Prior research suggests that novices would need simplified functionalities, whereas experts might prefer advanced features, customization, and enhanced visual control. In our survey, we noticed unexpected patterns between user profiles and flexibility scores. Most tools assumed expert users and provided limited flexibility, while a few were designed for novices, such as Zooids and ReACH offering highly flexible features. 
We observed insufficient involvement of decision makers in the design of decision-focused tools.
Even when decision makers were involved, it was usually post-initial prototyping or during the final evaluation. Consistent with prior reviews \cite{Dimara2021DecisionTasks}, few studies asked participants to undertake decision tasks. This lack of early involvement and minimal attention to decision makers and tasks during the requirement analysis phase can lead to poorly defined decision problems. To optimize tool effectiveness, early engagement with real decision makers is vital, delivering insights into needs, preferences, and practices and ensuring context alignment \cite{mahyar2020designing}. We suggest adopting comprehensive methods, including interviews, requirements, and task analysis for a thorough delineation of the decision making problem.}

\section{Limitations}
Our study has limitations that should be considered when interpreting our findings. Our \tag{visibility} metric is based on what the coders could independently agree and code, with the limited value range in our tags due to the lack of detailed  descriptions in some of the papers. Moreover, our \tag{flexibility} metric emphasizes the diversity of interactions across classes rather than the number of interactions within each class. These scores indicate the number of various features offered by each tool and should not be interpreted as negative or positive judgments.
Moreover, our study does not assess the real effectiveness of interactions or representations, and we hope our metrics, tags, and analysis can help design evaluations in future research. \revision{
Additionally, papers without the word \quotes{decision} in abstracts or titles might still be relevant tools \cite{Gratzl2013Lineup} or concepts like comparison \cite{Gleicher2017Comparison} for decision making.
Venues other than those we reviewed 
such as from the GIS \cite{malczewski2020emerging} or PubMed \cite{park2022impact} community, may also have pertinent papers.}

\section{Conclusion}

\revision{Our survey underscores a surprising scarcity of visualization tools
that support all stages of the decision making process.
We suggest future research augment decision making tools' flexibility,
involving exploration of novel paradigms like physical visualizations or immersive technologies, catering to diverse flexibility levels, enabling decision makers to navigate complex decisions.
Concurrently, designers could intertwine algorithmic support with user control via the enhanced flexibility levels, thereby aiding preference articulation and oversight. }
We hope that the metrics and review of decision sciences presented in this study serve as valuable resources for designing novel visualization tools.
Designers can leverage these metrics to scrutinize and enrich their designs with additional flexibility and visibility levels while addressing domain-specific challenges by considering decision alternatives and criteria preferences explicitly. 
Our study aims to empower all decision makers with effective visualization tools, from the everyday citizen to the highest levels of leadership, to make better-informed decisions and move seamlessly from \emph{information} to \emph{choice}.

\vspace{-0.5em}
\acknowledgments{
\vspace{-0.5em}We thank M. Jasim for his precious feedback and the authors of the papers in Fig. 5 who kindly and swiftly provided their own figures.}
\vspace{-0.5em}

\bibliographystyle{abbrv-doi-hyperref-narrow}

\bibliography{00_bibliography}

\begin{thebibliography}{100}
\renewcommand*{\sfdefault}{PTSansNarrow-TLF}

\bibitem{Afzal2011Epidemic}
S.~Afzal, R.~Maciejewski, and D.~S. Ebert.
\newblock Visual analytics decision support environment for epidemic modeling
  and response evaluation.
\newblock In {\em Proc. VAST}, pp. 191--200. IEEE, Piscataway, 2011.
  \href{https://doi.org/10.1109/VAST.2011.6102457}
{doi: \textsf{%
10\hspace{.1pt}\discretionary{.}{%
}{.}\hspace{.4pt}1109\discretionary{/}{%
}{/}VAST\hspace{.1pt}\discretionary{.}{%
}{.}\hspace{.4pt}2011\hspace{.1pt}\discretionary{.}{%
}{.}\hspace{.4pt}6102457}}


\bibitem{Ahn2019FairSight}
Y.~Ahn and Y.-R. Lin.
\newblock Fairsight: Visual analytics for fairness in decision making.
\newblock {\em IEEE TVCG}, 26(1):1086--1095, 2020.
  \href{https://doi.org/10.1109/TVCG.2019.2934262}
{doi: \textsf{%
10\hspace{.1pt}\discretionary{.}{%
}{.}\hspace{.4pt}1109\discretionary{/}{%
}{/}TVCG\hspace{.1pt}\discretionary{.}{%
}{.}\hspace{.4pt}2019\hspace{.1pt}\discretionary{.}{%
}{.}\hspace{.4pt}2934262}}


\bibitem{alves2023exploring}
T.~Alves, T.~Delgado, J.~Henriques-Calado, D.~Gon{\c{c}}alves, and S.~Gama.
\newblock Exploring the role of conscientiousness on visualization-supported
  decision-making.
\newblock {\em Comput. Graphics}, 111:47--62, 2023.
  \href{https://doi.org/10.1016/j.cag.2023.01.010}
{doi: \textsf{%
10\hspace{.1pt}\discretionary{.}{%
}{.}\hspace{.4pt}1016\discretionary{/}{%
}{/}j\hspace{.1pt}\discretionary{.}{%
}{.}\hspace{.4pt}cag\hspace{.1pt}\discretionary{.}{%
}{.}\hspace{.4pt}2023\hspace{.1pt}\discretionary{.}{%
}{.}\hspace{.4pt}01\hspace{.1pt}\discretionary{.}{%
}{.}\hspace{.4pt}010}}


\bibitem{Andrienko2007Evacuation}
G.~Andrienko, N.~Andrienko, and U.~Bartling.
\newblock Visual analytics approach to user-controlled evacuation scheduling.
\newblock In {\em Proc. VAST}, pp. 43--50. IEEE, 2007.
  \href{https://doi.org/10.1109/VAST.2007.4388995}
{doi: \textsf{%
10\hspace{.1pt}\discretionary{.}{%
}{.}\hspace{.4pt}1109\discretionary{/}{%
}{/}VAST\hspace{.1pt}\discretionary{.}{%
}{.}\hspace{.4pt}2007\hspace{.1pt}\discretionary{.}{%
}{.}\hspace{.4pt}4388995}}


\bibitem{Asahi1995Treemaps}
T.~Asahi, D.~Turo, and B.~Shneiderman.
\newblock Visual decision-making: using treemaps for the analytic hierarchy
  process.
\newblock In {\em Proc. CHI}, pp. 405--406. ACM, 1995.
  \href{https://doi.org/10.1145/223355.223747}
{doi: \textsf{%
10\hspace{.1pt}\discretionary{.}{%
}{.}\hspace{.4pt}1145\discretionary{/}{%
}{/}223355\hspace{.1pt}\discretionary{.}{%
}{.}\hspace{.4pt}223747}}


\bibitem{Aseniero2015Stratos}
B.~A. Aseniero, T.~Wun, D.~Ledo, G.~Ruhe, A.~Tang, and S.~Carpendale.
\newblock Stratos: Using visualization to support decisions in strategic
  software release planning.
\newblock In {\em Proc. CHI}, pp. 1479--1488. ACM, New York, 2015.
  \href{https://doi.org/10.1145/2702123.2702426}
{doi: \textsf{%
10\hspace{.1pt}\discretionary{.}{%
}{.}\hspace{.4pt}1145\discretionary{/}{%
}{/}2702123\hspace{.1pt}\discretionary{.}{%
}{.}\hspace{.4pt}2702426}}


\bibitem{azuma1999}
R.~Azuma, H.~Neely, M.~Daily, and M.~Correa.
\newblock Visualization of conflicts and resolutions in a" free flight"
  scenario.
\newblock In {\em Proc. VIS}, pp. 433--436. IEEE, Piscataway, 1999.
\newblock Color plate on page 557.
  \href{https://doi.org/10.1109/VISUAL.1999.809923}
{doi: \textsf{%
10\hspace{.1pt}\discretionary{.}{%
}{.}\hspace{.4pt}1109\discretionary{/}{%
}{/}VISUAL\hspace{.1pt}\discretionary{.}{%
}{.}\hspace{.4pt}1999\hspace{.1pt}\discretionary{.}{%
}{.}\hspace{.4pt}809923}}


\bibitem{bajracharya2018interactive}
S.~Bajracharya, G.~Carenini, B.~Chamberlain, K.~Chen, D.~Klein, D.~Poole,
  H.~Taheri, and G.~{\"O}berg.
\newblock Interactive visualization for group decision analysis.
\newblock {\em Int. J. Inf. Technol. Decis. Making}, 17(6):1839--1864, 2018.
  \href{https://doi.org/10.1142/S0219622018500384}
{doi: \textsf{%
10\hspace{.1pt}\discretionary{.}{%
}{.}\hspace{.4pt}1142\discretionary{/}{%
}{/}S0219622018500384}}


\bibitem{bartram2021untidy}
L.~Bartram, M.~Correll, and M.~Tory.
\newblock Untidy data: The unreasonable effectiveness of tables.
\newblock {\em IEEE TVCG}, 28(1):686--696, 2022.
  \href{https://doi.org/10.1109/TVCG.2021.3114830}
{doi: \textsf{%
10\hspace{.1pt}\discretionary{.}{%
}{.}\hspace{.4pt}1109\discretionary{/}{%
}{/}TVCG\hspace{.1pt}\discretionary{.}{%
}{.}\hspace{.4pt}2021\hspace{.1pt}\discretionary{.}{%
}{.}\hspace{.4pt}3114830}}


\bibitem{Baumer2022Political}
E.~P. Baumer, M.~Jasim, A.~Sarvghad, and N.~Mahyar.
\newblock Of course it's political! a critical inquiry into underemphasized
  dimensions in civic text visualization.
\newblock {\em Comput. Graphics Forum}, 41(3):1--14, 2022.
  \href{https://doi.org/10.1111/cgf.14518}
{doi: \textsf{%
10\hspace{.1pt}\discretionary{.}{%
}{.}\hspace{.4pt}1111\discretionary{/}{%
}{/}cgf\hspace{.1pt}\discretionary{.}{%
}{.}\hspace{.4pt}14518}}


\bibitem{Bautista2006Framework}
J.~Bautista and G.~Carenini.
\newblock An integrated task-based framework for the design and evaluation of
  visualizations to support preferential choice.
\newblock In {\em Proc. AVI}, pp. 217--224. ACM, 2006.
  \href{https://doi.org/10.1145/1133265.1133308}
{doi: \textsf{%
10\hspace{.1pt}\discretionary{.}{%
}{.}\hspace{.4pt}1145\discretionary{/}{%
}{/}1133265\hspace{.1pt}\discretionary{.}{%
}{.}\hspace{.4pt}1133308}}


\bibitem{Booshehrian2012Vismon}
M.~Booshehrian, T.~M{\"o}ller, R.~M. Peterman, and T.~Munzner.
\newblock Vismon: Facilitating analysis of trade-offs, uncertainty, and
  sensitivity in fisheries management decision making.
\newblock {\em Comput. Graphics Forum}, 31(3pt3):1235--1244, 2012.
  \href{https://doi.org/10.1111/j.1467-8659.2012.03116.x}
{doi: \textsf{%
10\hspace{.1pt}\discretionary{.}{%
}{.}\hspace{.4pt}1111\discretionary{/}{%
}{/}j\hspace{.1pt}\discretionary{.}{%
}{.}\hspace{.4pt}1467\discretionary{%
}{-}{-}8659\hspace{.1pt}\discretionary{.}{%
}{.}\hspace{.4pt}2012\hspace{.1pt}\discretionary{.}{%
}{.}\hspace{.4pt}03116\hspace{.1pt}\discretionary{.}{%
}{.}\hspace{.4pt}x}}


\bibitem{brehmer2021jam}
M.~Brehmer and R.~Kosara.
\newblock From jam session to recital: Synchronous communication and
  collaboration around data in organizations.
\newblock {\em IEEE TVCG}, 28(1):1139--1149, 2022.
  \href{https://doi.org/10.1109/TVCG.2021.3114760}
{doi: \textsf{%
10\hspace{.1pt}\discretionary{.}{%
}{.}\hspace{.4pt}1109\discretionary{/}{%
}{/}TVCG\hspace{.1pt}\discretionary{.}{%
}{.}\hspace{.4pt}2021\hspace{.1pt}\discretionary{.}{%
}{.}\hspace{.4pt}3114760}}


\bibitem{Cao2022TeamBuilder}
A.~Cao, J.~Lan, X.~Xie, H.~Chen, X.~Zhang, H.~Zhang, and Y.~Wu.
\newblock Team-builder: Toward more effective lineup selection in soccer.
\newblock {\em IEEE TVCG}, pp. 1--16, 2022.
\newblock Early Access. \href{https://doi.org/10.1109/TVCG.2022.3207147}
{doi: \textsf{%
10\hspace{.1pt}\discretionary{.}{%
}{.}\hspace{.4pt}1109\discretionary{/}{%
}{/}TVCG\hspace{.1pt}\discretionary{.}{%
}{.}\hspace{.4pt}2022\hspace{.1pt}\discretionary{.}{%
}{.}\hspace{.4pt}3207147}}


\bibitem{Carenini2004ValueCharts}
G.~Carenini and J.~Loyd.
\newblock Valuecharts: analyzing linear models expressing preferences and
  evaluations.
\newblock In {\em Proc. AVI}, pp. 150--157. ACM, New York, 2004.
  \href{https://doi.org/10.1145/989863.989885}
{doi: \textsf{%
10\hspace{.1pt}\discretionary{.}{%
}{.}\hspace{.4pt}1145\discretionary{/}{%
}{/}989863\hspace{.1pt}\discretionary{.}{%
}{.}\hspace{.4pt}989885}}


\bibitem{Castro2022Effort}
S.~C. Castro, P.~S. Quinan, H.~Hosseinpour, and L.~Padilla.
\newblock Examining effort in 1d uncertainty communication using individual
  differences in working memory and nasa-tlx.
\newblock {\em IEEE TVCG}, 28(1):411--421, 2022.
  \href{https://doi.org/10.1109/TVCG.2021.3114803}
{doi: \textsf{%
10\hspace{.1pt}\discretionary{.}{%
}{.}\hspace{.4pt}1109\discretionary{/}{%
}{/}TVCG\hspace{.1pt}\discretionary{.}{%
}{.}\hspace{.4pt}2021\hspace{.1pt}\discretionary{.}{%
}{.}\hspace{.4pt}3114803}}


\bibitem{Chatzimparmpas2020Trust}
A.~Chatzimparmpas, R.~M. Martins, I.~Jusufi, K.~Kucher, F.~Rossi, and
  A.~Kerren.
\newblock The state of the art in enhancing trust in machine learning models
  with the use of visualizations.
\newblock {\em Comput. Graphics Forum}, 39(3):713--756, 2020.
  \href{https://doi.org/10.1111/cgf.14034}
{doi: \textsf{%
10\hspace{.1pt}\discretionary{.}{%
}{.}\hspace{.4pt}1111\discretionary{/}{%
}{/}cgf\hspace{.1pt}\discretionary{.}{%
}{.}\hspace{.4pt}14034}}


\bibitem{chermack2011scenario}
T.~J. Chermack.
\newblock {\em Scenario planning in organizations: how to create, use, and
  assess scenarios}.
\newblock Berrett-Koehler Publishers, San Francisco, 2011.

\bibitem{Cibulski2020PAVED}
L.~Cibulski, H.~Mitterhofer, T.~May, and J.~Kohlhammer.
\newblock Paved: Pareto front visualization for engineering design.
\newblock {\em Comput. Graphics Forum}, 39(3):405--416, 2020.
  \href{https://doi.org/10.1111/cgf.13990}
{doi: \textsf{%
10\hspace{.1pt}\discretionary{.}{%
}{.}\hspace{.4pt}1111\discretionary{/}{%
}{/}cgf\hspace{.1pt}\discretionary{.}{%
}{.}\hspace{.4pt}13990}}


\bibitem{cohen}
J.~Cohen.
\newblock A coefficient of agreement for nominal scales.
\newblock {\em EPM}, 20(1):37--46, 1960.
  \href{https://doi.org/10.1177/001316446002000104}
{doi: \textsf{%
10\hspace{.1pt}\discretionary{.}{%
}{.}\hspace{.4pt}1177\discretionary{/}{%
}{/}001316446002000104}}


\bibitem{Dhurkari2022Mcdm}
R.~K. Dhurkari.
\newblock Mcdm methods: Practical difficulties and future directions for
  improvement.
\newblock {\em Rairo Oper. Res.}, 56(4):2221--2233, 2022.
  \href{https://doi.org/10.1051/ro/2022060}
{doi: \textsf{%
10\hspace{.1pt}\discretionary{.}{%
}{.}\hspace{.4pt}1051\discretionary{/}{%
}{/}ro\discretionary{/}{%
}{/}2022060}}


\bibitem{Dimara2019Mitigating}
E.~Dimara, G.~Bailly, A.~Bezerianos, and S.~Franconeri.
\newblock Mitigating the attraction effect with visualizations.
\newblock {\em IEEE TVCG}, 25(1):850--860, 2019.
  \href{https://doi.org/10.1109/TVCG.2018.2865233}
{doi: \textsf{%
10\hspace{.1pt}\discretionary{.}{%
}{.}\hspace{.4pt}1109\discretionary{/}{%
}{/}TVCG\hspace{.1pt}\discretionary{.}{%
}{.}\hspace{.4pt}2018\hspace{.1pt}\discretionary{.}{%
}{.}\hspace{.4pt}2865233}}


\bibitem{Dimara2018DecisionSupport}
E.~Dimara, A.~Bezerianos, and P.~Dragicevic.
\newblock Conceptual and methodological issues in evaluating multidimensional
  visualizations for decision support.
\newblock {\em IEEE TVCG}, 24(1):749--759, 2018.
  \href{https://doi.org/10.1109/TVCG.2017.2745138}
{doi: \textsf{%
10\hspace{.1pt}\discretionary{.}{%
}{.}\hspace{.4pt}1109\discretionary{/}{%
}{/}TVCG\hspace{.1pt}\discretionary{.}{%
}{.}\hspace{.4pt}2017\hspace{.1pt}\discretionary{.}{%
}{.}\hspace{.4pt}2745138}}


\bibitem{Dimara2020Interaction}
E.~Dimara and C.~Perin.
\newblock What is interaction for data visualization?
\newblock {\em IEEE TVCG}, 26(1):119--129, 2020.
  \href{https://doi.org/10.1109/TVCG.2019.2934283}
{doi: \textsf{%
10\hspace{.1pt}\discretionary{.}{%
}{.}\hspace{.4pt}1109\discretionary{/}{%
}{/}TVCG\hspace{.1pt}\discretionary{.}{%
}{.}\hspace{.4pt}2019\hspace{.1pt}\discretionary{.}{%
}{.}\hspace{.4pt}2934283}}


\bibitem{Dimara2021DecisionTasks}
E.~Dimara and J.~Stasko.
\newblock A critical reflection on visualization research: Where do decision
  making tasks hide?
\newblock {\em IEEE TVCG}, 28(1):1128--1138, 2022.
  \href{https://doi.org/10.1109/TVCG.2021.3114813}
{doi: \textsf{%
10\hspace{.1pt}\discretionary{.}{%
}{.}\hspace{.4pt}1109\discretionary{/}{%
}{/}TVCG\hspace{.1pt}\discretionary{.}{%
}{.}\hspace{.4pt}2021\hspace{.1pt}\discretionary{.}{%
}{.}\hspace{.4pt}3114813}}


\bibitem{Dimara2021Organizations}
E.~Dimara, H.~Zhang, M.~Tory, and S.~Franconeri.
\newblock The unmet data visualization needs of decision makers within
  organizations.
\newblock {\em IEEE TVCG}, 28(12):4101--4112, 2022.
  \href{https://doi.org/10.1109/TVCG.2021.3074023}
{doi: \textsf{%
10\hspace{.1pt}\discretionary{.}{%
}{.}\hspace{.4pt}1109\discretionary{/}{%
}{/}TVCG\hspace{.1pt}\discretionary{.}{%
}{.}\hspace{.4pt}2021\hspace{.1pt}\discretionary{.}{%
}{.}\hspace{.4pt}3074023}}


\bibitem{Dy2021Improving}
B.~Dy, N.~Ibrahim, A.~Poorthuis, and S.~Joyce.
\newblock Improving visualization design for effective multi-objective decision
  making.
\newblock {\em IEEE TVCG}, 28(10):3405--3416, 2022.
  \href{https://doi.org/10.1109/TVCG.2021.3065126}
{doi: \textsf{%
10\hspace{.1pt}\discretionary{.}{%
}{.}\hspace{.4pt}1109\discretionary{/}{%
}{/}TVCG\hspace{.1pt}\discretionary{.}{%
}{.}\hspace{.4pt}2021\hspace{.1pt}\discretionary{.}{%
}{.}\hspace{.4pt}3065126}}


\bibitem{elo2008qualitative}
S.~Elo and H.~Kyng{\"a}s.
\newblock The qualitative content analysis process.
\newblock {\em JAN}, 62(1):107--115, 2008.
  \href{https://doi.org/10.1111/j.1365-2648.2007.04569.x}
{doi: \textsf{%
10\hspace{.1pt}\discretionary{.}{%
}{.}\hspace{.4pt}1111\discretionary{/}{%
}{/}j\hspace{.1pt}\discretionary{.}{%
}{.}\hspace{.4pt}1365\discretionary{%
}{-}{-}2648\hspace{.1pt}\discretionary{.}{%
}{.}\hspace{.4pt}2007\hspace{.1pt}\discretionary{.}{%
}{.}\hspace{.4pt}04569\hspace{.1pt}\discretionary{.}{%
}{.}\hspace{.4pt}x}}


\bibitem{Feng2020Topology}
Z.~Feng, H.~Li, W.~Zeng, S.-H. Yang, and H.~Qu.
\newblock Topology density map for urban data visualization and analysis.
\newblock {\em IEEE TVCG}, 27(2):828--838, 2021.
  \href{https://doi.org/10.1109/TVCG.2020.3030469}
{doi: \textsf{%
10\hspace{.1pt}\discretionary{.}{%
}{.}\hspace{.4pt}1109\discretionary{/}{%
}{/}TVCG\hspace{.1pt}\discretionary{.}{%
}{.}\hspace{.4pt}2020\hspace{.1pt}\discretionary{.}{%
}{.}\hspace{.4pt}3030469}}


\bibitem{Ferreira2015Urbane}
N.~Ferreira, M.~Lage, H.~Doraiswamy, H.~Vo, L.~Wilson, H.~Werner, M.~Park, and
  C.~Silva.
\newblock Urbane: A 3d framework to support data driven decision making in
  urban development.
\newblock In {\em Proc. VAST}, pp. 97--104. IEEE, Piscataway, 2015.
  \href{https://doi.org/10.1109/VAST.2015.7347636}
{doi: \textsf{%
10\hspace{.1pt}\discretionary{.}{%
}{.}\hspace{.4pt}1109\discretionary{/}{%
}{/}VAST\hspace{.1pt}\discretionary{.}{%
}{.}\hspace{.4pt}2015\hspace{.1pt}\discretionary{.}{%
}{.}\hspace{.4pt}7347636}}


\bibitem{Greco2016MCDA}
J.~Figueira, S.~Greco, and M.~Ehrgott.
\newblock {\em Multiple criteria decision analysis: State of the art surveys}.
\newblock Springer, New York, 2005. \href{https://doi.org/10.1007/b100605}
{doi: \textsf{%
10\hspace{.1pt}\discretionary{.}{%
}{.}\hspace{.4pt}1007\discretionary{/}{%
}{/}b100605}}


\bibitem{statistical}
J.~L. Fleiss, B.~Levin, and M.~C. Paik.
\newblock {\em Statistical methods for rates and proportions}.
\newblock John Wiley \& Sons, Inc., New Jersey, 2003.
  \href{https://doi.org/10.1002/0471445428}
{doi: \textsf{%
10\hspace{.1pt}\discretionary{.}{%
}{.}\hspace{.4pt}1002\discretionary{/}{%
}{/}0471445428}}


\bibitem{gettinger2013comparison}
J.~Gettinger, E.~Kiesling, C.~Stummer, and R.~Vetschera.
\newblock A comparison of representations for discrete multi-criteria decision
  problems.
\newblock {\em Decis. Support Syst.}, 54(2):976--985, 2013.
  \href{https://doi.org/10.1016/j.dss.2012.10.023}
{doi: \textsf{%
10\hspace{.1pt}\discretionary{.}{%
}{.}\hspace{.4pt}1016\discretionary{/}{%
}{/}j\hspace{.1pt}\discretionary{.}{%
}{.}\hspace{.4pt}dss\hspace{.1pt}\discretionary{.}{%
}{.}\hspace{.4pt}2012\hspace{.1pt}\discretionary{.}{%
}{.}\hspace{.4pt}10\hspace{.1pt}\discretionary{.}{%
}{.}\hspace{.4pt}023}}


\bibitem{Gleicher2017Comparison}
M.~Gleicher.
\newblock Considerations for visualizing comparison.
\newblock {\em IEEE TVCG}, 24(1):413--423, 2018.
  \href{https://doi.org/10.1109/TVCG.2017.2744199}
{doi: \textsf{%
10\hspace{.1pt}\discretionary{.}{%
}{.}\hspace{.4pt}1109\discretionary{/}{%
}{/}TVCG\hspace{.1pt}\discretionary{.}{%
}{.}\hspace{.4pt}2017\hspace{.1pt}\discretionary{.}{%
}{.}\hspace{.4pt}2744199}}


\bibitem{Gratzl2013Lineup}
S.~Gratzl, A.~Lex, N.~Gehlenborg, H.~Pfister, and M.~Streit.
\newblock Lineup: Visual analysis of multi-attribute rankings.
\newblock {\em IEEE TVCG}, 19(12):2277--2286, 2013.
  \href{https://doi.org/10.1109/TVCG.2013.173}
{doi: \textsf{%
10\hspace{.1pt}\discretionary{.}{%
}{.}\hspace{.4pt}1109\discretionary{/}{%
}{/}TVCG\hspace{.1pt}\discretionary{.}{%
}{.}\hspace{.4pt}2013\hspace{.1pt}\discretionary{.}{%
}{.}\hspace{.4pt}173}}


\bibitem{greening2004design}
L.~A. Greening and S.~Bernow.
\newblock Design of coordinated energy and environmental policies: use of
  multi-criteria decision-making.
\newblock {\em Energy Policy}, 32(6):721--735, 2004.
  \href{https://doi.org/10.1016/j.enpol.2003.08.017}
{doi: \textsf{%
10\hspace{.1pt}\discretionary{.}{%
}{.}\hspace{.4pt}1016\discretionary{/}{%
}{/}j\hspace{.1pt}\discretionary{.}{%
}{.}\hspace{.4pt}enpol\hspace{.1pt}\discretionary{.}{%
}{.}\hspace{.4pt}2003\hspace{.1pt}\discretionary{.}{%
}{.}\hspace{.4pt}08\hspace{.1pt}\discretionary{.}{%
}{.}\hspace{.4pt}017}}


\bibitem{Guo2019Event}
S.~Guo, F.~Du, S.~Malik, E.~Koh, S.~Kim, Z.~Liu, D.~Kim, H.~Zha, and N.~Cao.
\newblock Visualizing uncertainty and alternatives in event sequence
  predictions.
\newblock In {\em Proc. CHI}, pp. 573:1--573:12. ACM, New York, 2019.
  \href{https://doi.org/10.1145/3290605.3300803}
{doi: \textsf{%
10\hspace{.1pt}\discretionary{.}{%
}{.}\hspace{.4pt}1145\discretionary{/}{%
}{/}3290605\hspace{.1pt}\discretionary{.}{%
}{.}\hspace{.4pt}3300803}}


\bibitem{ham2009integrating}
C.~Ham and J.~Oldham.
\newblock Integrating health and social care in england: lessons from early
  adopters and implications for policy.
\newblock {\em JICA}, 17(6):3--9, 2009.
  \href{https://doi.org/10.1108/14769018200900040}
{doi: \textsf{%
10\hspace{.1pt}\discretionary{.}{%
}{.}\hspace{.4pt}1108\discretionary{/}{%
}{/}14769018200900040}}


\bibitem{han2022kicking}
Y.~Han.
\newblock Kicking analysts out of the meeting room: Supporting future
  data-driven decision making with intelligent interactive visualization
  systems.
\newblock In {\em Proc. TREX}, pp. 16--21. IEEE, Piscataway, 2022.
  \href{https://doi.org/10.1109/TREX57753.2022.00007}
{doi: \textsf{%
10\hspace{.1pt}\discretionary{.}{%
}{.}\hspace{.4pt}1109\discretionary{/}{%
}{/}TREX57753\hspace{.1pt}\discretionary{.}{%
}{.}\hspace{.4pt}2022\hspace{.1pt}\discretionary{.}{%
}{.}\hspace{.4pt}00007}}


\bibitem{Handler2022ClioQuery}
A.~Handler, N.~Mahyar, and B.~O’Connor.
\newblock Clioquery: Interactive query-oriented text analytics for
  comprehensive investigation of historical news archives.
\newblock {\em ACM TiiS}, 12(3):22:1--22:49, 2022.
  \href{https://doi.org/10.1145/3524025}
{doi: \textsf{%
10\hspace{.1pt}\discretionary{.}{%
}{.}\hspace{.4pt}1145\discretionary{/}{%
}{/}3524025}}


\bibitem{Hindalong2022GroupDecisions}
E.~Hindalong, J.~Johnson, G.~Carenini, and T.~Munzner.
\newblock Abstractions for visualizing preferences in group decisions.
\newblock {\em ACM HCI}, 6(CSCW1):49:1--49:44, 2022.
  \href{https://doi.org/10.1145/3512896}
{doi: \textsf{%
10\hspace{.1pt}\discretionary{.}{%
}{.}\hspace{.4pt}1145\discretionary{/}{%
}{/}3512896}}


\bibitem{Huron2021DataInput}
S.~Huron and W.~Willett.
\newblock Visualizations as data input?
\newblock In {\em Proc. VIS}. IEEE, 2021.
\newblock presented at alt.VIS workshop,
  \url{https://altvis.github.io/2021.html}.

\bibitem{jansen2013interaction}
Y.~Jansen and P.~Dragicevic.
\newblock An interaction model for visualizations beyond the desktop.
\newblock {\em IEEE TVCG}, 19(12):2396--2405, 2013.
  \href{https://doi.org/10.1109/TVCG.2013.134}
{doi: \textsf{%
10\hspace{.1pt}\discretionary{.}{%
}{.}\hspace{.4pt}1109\discretionary{/}{%
}{/}TVCG\hspace{.1pt}\discretionary{.}{%
}{.}\hspace{.4pt}2013\hspace{.1pt}\discretionary{.}{%
}{.}\hspace{.4pt}134}}


\bibitem{Jasim2022Supporting}
M.~Jasim, C.~Collins, A.~Sarvghad, and N.~Mahyar.
\newblock Supporting serendipitous discovery and balanced analysis of online
  product reviews with interaction-driven metrics and bias-mitigating
  suggestions.
\newblock In {\em Proc. CHI}, pp. 9:1--9:24. ACM, New York, 2022.
  \href{https://doi.org/10.1145/3491102.3517649}
{doi: \textsf{%
10\hspace{.1pt}\discretionary{.}{%
}{.}\hspace{.4pt}1145\discretionary{/}{%
}{/}3491102\hspace{.1pt}\discretionary{.}{%
}{.}\hspace{.4pt}3517649}}


\bibitem{Jasim2021CommunityPulse}
M.~Jasim, E.~Hoque, A.~Sarvghad, and N.~Mahyar.
\newblock Communitypulse: Facilitating community input analysis by surfacing
  hidden insights, reflections, and priorities.
\newblock In {\em Proc. DIS}, pp. 846--863. ACM, New York, 2021.
  \href{https://doi.org/10.1145/3461778.3462132}
{doi: \textsf{%
10\hspace{.1pt}\discretionary{.}{%
}{.}\hspace{.4pt}1145\discretionary{/}{%
}{/}3461778\hspace{.1pt}\discretionary{.}{%
}{.}\hspace{.4pt}3462132}}


\bibitem{Kale2020VisualReasoning}
A.~Kale, M.~Kay, and J.~Hullman.
\newblock Visual reasoning strategies for effect size judgments and decisions.
\newblock {\em IEEE TVCG}, 27(2):272--282, 2021.
  \href{https://doi.org/10.1109/TVCG.2020.3030335}
{doi: \textsf{%
10\hspace{.1pt}\discretionary{.}{%
}{.}\hspace{.4pt}1109\discretionary{/}{%
}{/}TVCG\hspace{.1pt}\discretionary{.}{%
}{.}\hspace{.4pt}2020\hspace{.1pt}\discretionary{.}{%
}{.}\hspace{.4pt}3030335}}


\bibitem{Kandogan2014Insight}
E.~Kandogan, A.~Balakrishnan, E.~M. Haber, and J.~S. Pierce.
\newblock From data to insight: work practices of analysts in the enterprise.
\newblock {\em IEEE CG\&A}, 34(5):42--50, 2014.
  \href{https://doi.org/10.1109/MCG.2014.62}
{doi: \textsf{%
10\hspace{.1pt}\discretionary{.}{%
}{.}\hspace{.4pt}1109\discretionary{/}{%
}{/}MCG\hspace{.1pt}\discretionary{.}{%
}{.}\hspace{.4pt}2014\hspace{.1pt}\discretionary{.}{%
}{.}\hspace{.4pt}62}}


\bibitem{Keefe2010Workflows}
D.~Keefe.
\newblock {Integrating visualization and interaction research to improve
  scientific workflows}.
\newblock {\em IEEE CG\&A}, 30(2):8--13, 2010.
  \href{https://doi.org/10.1109/MCG.2010.30}
{doi: \textsf{%
10\hspace{.1pt}\discretionary{.}{%
}{.}\hspace{.4pt}1109\discretionary{/}{%
}{/}MCG\hspace{.1pt}\discretionary{.}{%
}{.}\hspace{.4pt}2010\hspace{.1pt}\discretionary{.}{%
}{.}\hspace{.4pt}30}}


\bibitem{keeney1982decision}
R.~L. Keeney.
\newblock Decision analysis: an overview.
\newblock {\em Oper. Res.}, 30(5):803--838, 1982.
  \href{https://doi.org/10.1287/opre.30.5.803}
{doi: \textsf{%
10\hspace{.1pt}\discretionary{.}{%
}{.}\hspace{.4pt}1287\discretionary{/}{%
}{/}opre\hspace{.1pt}\discretionary{.}{%
}{.}\hspace{.4pt}30\hspace{.1pt}\discretionary{.}{%
}{.}\hspace{.4pt}5\hspace{.1pt}\discretionary{.}{%
}{.}\hspace{.4pt}803}}


\bibitem{Keeney1993Decisions}
R.~L. Keeney and H.~Raiffa.
\newblock {\em Decisions with multiple objectives: preferences and value
  trade-offs}.
\newblock Cambridge University Press, Cambridge, 1993.

\bibitem{klein2008naturalistic}
G.~Klein.
\newblock Naturalistic decision making.
\newblock {\em Hum. Factors}, 50(3):456--460, 2008.
  \href{https://doi.org/10.1518/001872008X288385}
{doi: \textsf{%
10\hspace{.1pt}\discretionary{.}{%
}{.}\hspace{.4pt}1518\discretionary{/}{%
}{/}001872008X288385}}


\bibitem{kosara2023notebooks}
R.~Kosara.
\newblock Notebooks for data analysis and visualization: Moving beyond the
  data.
\newblock {\em IEEE CG\&A}, 43(1):91--96, 2023.
  \href{https://doi.org/10.1109/MCG.2022.3222024}
{doi: \textsf{%
10\hspace{.1pt}\discretionary{.}{%
}{.}\hspace{.4pt}1109\discretionary{/}{%
}{/}MCG\hspace{.1pt}\discretionary{.}{%
}{.}\hspace{.4pt}2022\hspace{.1pt}\discretionary{.}{%
}{.}\hspace{.4pt}3222024}}


\bibitem{Kreiser2018DecisionGraph}
J.~Kreiser, A.~Hann, E.~Zizer, and T.~Ropinski.
\newblock Decision graph embedding for high-resolution manometry diagnosis.
\newblock {\em IEEE TVCG}, 24(1):873--882, 2018.
  \href{https://doi.org/10.1109/TVCG.2017.2744299}
{doi: \textsf{%
10\hspace{.1pt}\discretionary{.}{%
}{.}\hspace{.4pt}1109\discretionary{/}{%
}{/}TVCG\hspace{.1pt}\discretionary{.}{%
}{.}\hspace{.4pt}2017\hspace{.1pt}\discretionary{.}{%
}{.}\hspace{.4pt}2744299}}


\bibitem{LeGoc2019Dynamic}
M.~Le~Goc, C.~Perin, S.~Follmer, J.-D. Fekete, and P.~Dragicevic.
\newblock Dynamic composite data physicalization using wheeled micro-robots.
\newblock {\em IEEE TVCG}, 25(1):737--747, 2019.
  \href{https://doi.org/10.1109/TVCG.2018.2865159}
{doi: \textsf{%
10\hspace{.1pt}\discretionary{.}{%
}{.}\hspace{.4pt}1109\discretionary{/}{%
}{/}TVCG\hspace{.1pt}\discretionary{.}{%
}{.}\hspace{.4pt}2018\hspace{.1pt}\discretionary{.}{%
}{.}\hspace{.4pt}2865159}}


\bibitem{malczewski2020emerging}
N.~J. Lim, S.~A. Brandt, and S.~Seipel.
\newblock Visualisation and evaluation of flood uncertainties based on ensemble
  modelling.
\newblock {\em Int. J. Geogr. Inf. Sci.}, 30(2):240--262, 2016.
  \href{https://doi.org/10.1080/13658816.2015.1085539}
{doi: \textsf{%
10\hspace{.1pt}\discretionary{.}{%
}{.}\hspace{.4pt}1080\discretionary{/}{%
}{/}13658816\hspace{.1pt}\discretionary{.}{%
}{.}\hspace{.4pt}2015\hspace{.1pt}\discretionary{.}{%
}{.}\hspace{.4pt}1085539}}


\bibitem{DataHunches2023}
H.~Lin, D.~Akbaba, M.~Meyer, and A.~Lex.
\newblock Data hunches: Incorporating personal knowledge into visualizations.
\newblock {\em IEEE TVCG}, 29(1):504--514, 2023.
  \href{https://doi.org/10.1109/TVCG.2022.3209451}
{doi: \textsf{%
10\hspace{.1pt}\discretionary{.}{%
}{.}\hspace{.4pt}1109\discretionary{/}{%
}{/}TVCG\hspace{.1pt}\discretionary{.}{%
}{.}\hspace{.4pt}2022\hspace{.1pt}\discretionary{.}{%
}{.}\hspace{.4pt}3209451}}


\bibitem{Liu2016Smartadp}
D.~Liu, D.~Weng, Y.~Li, J.~Bao, Y.~Zheng, H.~Qu, and Y.~Wu.
\newblock Smartadp: Visual analytics of large-scale taxi trajectories for
  selecting billboard locations.
\newblock {\em IEEE TVCG}, 23(1):1--10, 2017.
  \href{https://doi.org/10.1109/TVCG.2016.2598432}
{doi: \textsf{%
10\hspace{.1pt}\discretionary{.}{%
}{.}\hspace{.4pt}1109\discretionary{/}{%
}{/}TVCG\hspace{.1pt}\discretionary{.}{%
}{.}\hspace{.4pt}2016\hspace{.1pt}\discretionary{.}{%
}{.}\hspace{.4pt}2598432}}


\bibitem{Liu2010Mental}
Z.~Liu and J.~Stasko.
\newblock Mental models, visual reasoning and interaction in information
  visualization: A top-down perspective.
\newblock {\em IEEE TVCG}, 16(6):999--1008, 2010.
  \href{https://doi.org/10.1109/TVCG.2010.177}
{doi: \textsf{%
10\hspace{.1pt}\discretionary{.}{%
}{.}\hspace{.4pt}1109\discretionary{/}{%
}{/}TVCG\hspace{.1pt}\discretionary{.}{%
}{.}\hspace{.4pt}2010\hspace{.1pt}\discretionary{.}{%
}{.}\hspace{.4pt}177}}


\bibitem{macqueen1998codebook}
K.~M. MacQueen, E.~McLellan, K.~Kay, and B.~Milstein.
\newblock Codebook development for team-based qualitative analysis.
\newblock {\em Field Methods}, 10(2):31--36, 1998.
  \href{https://doi.org/10.1177/1525822X980100020301}
{doi: \textsf{%
10\hspace{.1pt}\discretionary{.}{%
}{.}\hspace{.4pt}1177\discretionary{/}{%
}{/}1525822X980100020301}}


\bibitem{mahyar2020designing}
N.~Mahyar, M.~Jasim, and A.~Sarvghad.
\newblock Designing technology for sociotechnical problems: challenges and
  considerations.
\newblock {\em IEEE CG\&A}, 40(6):76--87, 2020.
  \href{https://doi.org/10.1109/MCG.2020.3017405}
{doi: \textsf{%
10\hspace{.1pt}\discretionary{.}{%
}{.}\hspace{.4pt}1109\discretionary{/}{%
}{/}MCG\hspace{.1pt}\discretionary{.}{%
}{.}\hspace{.4pt}2020\hspace{.1pt}\discretionary{.}{%
}{.}\hspace{.4pt}3017405}}


\bibitem{Mahyar2017ConsesnsUs}
N.~Mahyar, W.~Liu, S.~Xiao, J.~Browne, M.~Yang, and S.~P. Dow.
\newblock Consesnsus: Visualizing points of disagreement for multi-criteria
  collaborative decision making.
\newblock In {\em Proc. CSCW}, pp. 17--20. ACM, New York, 2017.
  \href{https://doi.org/10.1145/3022198.3023269}
{doi: \textsf{%
10\hspace{.1pt}\discretionary{.}{%
}{.}\hspace{.4pt}1145\discretionary{/}{%
}{/}3022198\hspace{.1pt}\discretionary{.}{%
}{.}\hspace{.4pt}3023269}}


\bibitem{mahyar2019civic}
N.~Mahyar, D.~V. Nguyen, M.~Chan, J.~Zheng, and S.~P. Dow.
\newblock The civic data deluge: Understanding the challenges of analyzing
  large-scale community input.
\newblock In {\em Proc. DIS}, pp. 1171--1181. ACM, New York, 2019.
  \href{https://doi.org/10.1145/3322276.3322354}
{doi: \textsf{%
10\hspace{.1pt}\discretionary{.}{%
}{.}\hspace{.4pt}1145\discretionary{/}{%
}{/}3322276\hspace{.1pt}\discretionary{.}{%
}{.}\hspace{.4pt}3322354}}


\bibitem{Marriott2011HiTrees}
K.~Marriott, P.~Sbarski, T.~van Gelder, D.~Prager, and A.~Bulka.
\newblock Hi-trees and their layout.
\newblock {\em IEEE TVCG}, 17(3):290--304, 2011.
  \href{https://doi.org/10.1109/TVCG.2010.45}
{doi: \textsf{%
10\hspace{.1pt}\discretionary{.}{%
}{.}\hspace{.4pt}1109\discretionary{/}{%
}{/}TVCG\hspace{.1pt}\discretionary{.}{%
}{.}\hspace{.4pt}2010\hspace{.1pt}\discretionary{.}{%
}{.}\hspace{.4pt}45}}


\bibitem{mccurdy2018framework}
N.~McCurdy, J.~Gerdes, and M.~Meyer.
\newblock A framework for externalizing implicit error using visualization.
\newblock {\em IEEE TVCG}, 25(1):925--935, 2019.
  \href{https://doi.org/10.1109/TVCG.2018.2864913}
{doi: \textsf{%
10\hspace{.1pt}\discretionary{.}{%
}{.}\hspace{.4pt}1109\discretionary{/}{%
}{/}TVCG\hspace{.1pt}\discretionary{.}{%
}{.}\hspace{.4pt}2018\hspace{.1pt}\discretionary{.}{%
}{.}\hspace{.4pt}2864913}}


\bibitem{muller2021}
J.~M{\"u}ller, M.~Cypko, A.~Oeser, M.~Stoehr, V.~Zebralla, S.~Schreiber,
  S.~Wiegand, A.~Dietz, and S.~Oeltze-Jafra.
\newblock Visual assistance in clinical decision support.
\newblock In {\em Proc. EuroVis}. Eurographics, Geneva, 2021.
  \href{https://doi.org/10.2312/evm.20211075}
{doi: \textsf{%
10\hspace{.1pt}\discretionary{.}{%
}{.}\hspace{.4pt}2312\discretionary{/}{%
}{/}evm\hspace{.1pt}\discretionary{.}{%
}{.}\hspace{.4pt}20211075}}


\bibitem{nielsen2005ten}
J.~Nielsen.
\newblock Ten usability heuristics.
\newblock Website, 2020.
\newblock Accessed: 30 June 2023,
  \url{https://www.nngroup.com/articles/ten-usability-heuristics/}.

\bibitem{norman2013design}
D.~Norman.
\newblock {\em The design of everyday things: Revised and expanded edition}.
\newblock Basic Books, New York, 2013.

\bibitem{Pajer2017Weightlifter}
S.~Pajer, M.~Streit, T.~Torsney-Weir, F.~Spechtenhauser, T.~M{\"o}ller, and
  H.~Piringer.
\newblock Weightlifter: Visual weight space exploration for multi-criteria
  decision making.
\newblock {\em IEEE TVCG}, 23(1):611--620, 2017.
  \href{https://doi.org/10.1109/TVCG.2016.2598589}
{doi: \textsf{%
10\hspace{.1pt}\discretionary{.}{%
}{.}\hspace{.4pt}1109\discretionary{/}{%
}{/}TVCG\hspace{.1pt}\discretionary{.}{%
}{.}\hspace{.4pt}2016\hspace{.1pt}\discretionary{.}{%
}{.}\hspace{.4pt}2598589}}


\bibitem{park2022impact}
S.~Park, B.~Bekemeier, A.~Flaxman, and M.~Schultz.
\newblock Impact of data visualization on decision-making and its implications
  for public health practice: a systematic literature review.
\newblock {\em Inform. Health Soc. Care}, 47(2):175--193, 2022.
  \href{https://doi.org/10.1080/17538157.2021.1982949}
{doi: \textsf{%
10\hspace{.1pt}\discretionary{.}{%
}{.}\hspace{.4pt}1080\discretionary{/}{%
}{/}17538157\hspace{.1pt}\discretionary{.}{%
}{.}\hspace{.4pt}2021\hspace{.1pt}\discretionary{.}{%
}{.}\hspace{.4pt}1982949}}


\bibitem{payne2004walking}
J.~W. Payne and J.~R. Bettman.
\newblock Walking with the scarecrow: The information-processing approach to
  decision research.
\newblock In {\em Blackwell Handbook of Judgment and Decision Making}, chap.~6,
  pp. 110--132. John Wiley\& Sons, Inc, Newark, 2004.
  \href{https://doi.org/10.1002/9780470752937.ch6}
{doi: \textsf{%
10\hspace{.1pt}\discretionary{.}{%
}{.}\hspace{.4pt}1002\discretionary{/}{%
}{/}9780470752937\hspace{.1pt}\discretionary{.}{%
}{.}\hspace{.4pt}ch6}}


\bibitem{payne1988adaptive}
J.~W. Payne, J.~R. Bettman, and E.~J. Johnson.
\newblock Adaptive strategy selection in decision making.
\newblock {\em JEP:LMC}, 14(3):534--552, 1988.
  \href{https://doi.org/10.1037/0278-7393.14.3.534}
{doi: \textsf{%
10\hspace{.1pt}\discretionary{.}{%
}{.}\hspace{.4pt}1037\discretionary{/}{%
}{/}0278\discretionary{%
}{-}{-}7393\hspace{.1pt}\discretionary{.}{%
}{.}\hspace{.4pt}14\hspace{.1pt}\discretionary{.}{%
}{.}\hspace{.4pt}3\hspace{.1pt}\discretionary{.}{%
}{.}\hspace{.4pt}534}}


\bibitem{Pu2000SmartClient}
P.~Pu and B.~Faltings.
\newblock Enriching buyers' experiences: the smartclient approach.
\newblock In {\em Proc. CHI}, pp. 289--296. ACM, New York, 2000.
  \href{https://doi.org/10.1145/332040.332446}
{doi: \textsf{%
10\hspace{.1pt}\discretionary{.}{%
}{.}\hspace{.4pt}1145\discretionary{/}{%
}{/}332040\hspace{.1pt}\discretionary{.}{%
}{.}\hspace{.4pt}332446}}


\bibitem{reyna2011dual}
V.~F. Reyna and C.~J. Brainerd.
\newblock Dual processes in decision making and developmental neuroscience: A
  fuzzy-trace model.
\newblock {\em Dev. Rev.}, 31(2-3):180--206, 2011.
  \href{https://doi.org/10.1016/j.dr.2011.07.004}
{doi: \textsf{%
10\hspace{.1pt}\discretionary{.}{%
}{.}\hspace{.4pt}1016\discretionary{/}{%
}{/}j\hspace{.1pt}\discretionary{.}{%
}{.}\hspace{.4pt}dr\hspace{.1pt}\discretionary{.}{%
}{.}\hspace{.4pt}2011\hspace{.1pt}\discretionary{.}{%
}{.}\hspace{.4pt}07\hspace{.1pt}\discretionary{.}{%
}{.}\hspace{.4pt}004}}


\bibitem{Ribicic2012Sketching}
H.~Ribicic, J.~Waser, R.~Gurbat, B.~Sadransky, and M.~E. Gr{\"o}ller.
\newblock Sketching uncertainty into simulations.
\newblock {\em IEEE TVCG}, 18(12):2255--2264, 2012.
  \href{https://doi.org/10.1109/TVCG.2012.261}
{doi: \textsf{%
10\hspace{.1pt}\discretionary{.}{%
}{.}\hspace{.4pt}1109\discretionary{/}{%
}{/}TVCG\hspace{.1pt}\discretionary{.}{%
}{.}\hspace{.4pt}2012\hspace{.1pt}\discretionary{.}{%
}{.}\hspace{.4pt}261}}


\bibitem{Rudolph2009Finvis}
S.~Rudolph, A.~Savikhin, and D.~S. Ebert.
\newblock Finvis: Applied visual analytics for personal financial planning.
\newblock In {\em Proc. VAST}, pp. 195--202. IEEE, Piscataway, 2009.
  \href{https://doi.org/10.1109/VAST.2009.5333920}
{doi: \textsf{%
10\hspace{.1pt}\discretionary{.}{%
}{.}\hspace{.4pt}1109\discretionary{/}{%
}{/}VAST\hspace{.1pt}\discretionary{.}{%
}{.}\hspace{.4pt}2009\hspace{.1pt}\discretionary{.}{%
}{.}\hspace{.4pt}5333920}}


\bibitem{russo1983strategies}
J.~E. Russo and B.~A. Dosher.
\newblock Strategies for multiattribute binary choice.
\newblock {\em JEP:LMC}, 9(4):676--696, 1983.
  \href{https://doi.org/10.1037/0278-7393.9.4.676}
{doi: \textsf{%
10\hspace{.1pt}\discretionary{.}{%
}{.}\hspace{.4pt}1037\discretionary{/}{%
}{/}0278\discretionary{%
}{-}{-}7393\hspace{.1pt}\discretionary{.}{%
}{.}\hspace{.4pt}9\hspace{.1pt}\discretionary{.}{%
}{.}\hspace{.4pt}4\hspace{.1pt}\discretionary{.}{%
}{.}\hspace{.4pt}676}}


\bibitem{Savikhin2011FinancialPortfolio}
A.~Savikhin, H.~C. Lam, B.~Fisher, and D.~S. Ebert.
\newblock An experimental study of financial portfolio selection with visual
  analytics for decision support.
\newblock In {\em Proc. HICSS}, pp. 1--10. IEEE, Piscataway, 2011.
  \href{https://doi.org/10.1109/HICSS.2011.54}
{doi: \textsf{%
10\hspace{.1pt}\discretionary{.}{%
}{.}\hspace{.4pt}1109\discretionary{/}{%
}{/}HICSS\hspace{.1pt}\discretionary{.}{%
}{.}\hspace{.4pt}2011\hspace{.1pt}\discretionary{.}{%
}{.}\hspace{.4pt}54}}


\bibitem{siirtola2014bars}
H.~Siirtola.
\newblock Bars, pies, doughnuts \& tables--visualization of proportions.
\newblock In {\em Proc. BCS-HCI}, pp. 240--245. BCS, Swindon, 2014.
  \href{https://doi.org/10.14236/ewic/hci2014.30}
{doi: \textsf{%
10\hspace{.1pt}\discretionary{.}{%
}{.}\hspace{.4pt}14236\discretionary{/}{%
}{/}ewic\discretionary{/}{%
}{/}hci2014\hspace{.1pt}\discretionary{.}{%
}{.}\hspace{.4pt}30}}


\bibitem{Simon1960ManagementDecision}
H.~A. Simon.
\newblock {\em The new science of management decision.}
\newblock Harper \& Brothers, New York, 1960.
  \href{https://doi.org/10.1037/13978-000}
{doi: \textsf{%
10\hspace{.1pt}\discretionary{.}{%
}{.}\hspace{.4pt}1037\discretionary{/}{%
}{/}13978\discretionary{%
}{-}{-}000}}


\bibitem{Sorger2016Litevis}
J.~Sorger, T.~Ortner, C.~Luksch, M.~Schw{\"a}rzler, E.~Gr{\"o}ller, and
  H.~Piringer.
\newblock Litevis: integrated visualization for simulation-based decision
  support in lighting design.
\newblock {\em IEEE TVCG}, 22(1):290--299, 2016.
  \href{https://doi.org/10.1109/TVCG.2015.2468011}
{doi: \textsf{%
10\hspace{.1pt}\discretionary{.}{%
}{.}\hspace{.4pt}1109\discretionary{/}{%
}{/}TVCG\hspace{.1pt}\discretionary{.}{%
}{.}\hspace{.4pt}2015\hspace{.1pt}\discretionary{.}{%
}{.}\hspace{.4pt}2468011}}


\bibitem{Spence1998AttributeExplorer}
R.~Spence and L.~Tweedie.
\newblock The attribute explorer: information synthesis via exploration.
\newblock {\em Interact. Comput.}, 11(2):137--146, 1998.
  \href{https://doi.org/10.1016/S0953-5438(98)00022-8}
{doi: \textsf{%
10\hspace{.1pt}\discretionary{.}{%
}{.}\hspace{.4pt}1016\discretionary{/}{%
}{/}S0953\discretionary{%
}{-}{-}5438\discretionary{%
}{(}{(}98\discretionary{)}{%
}{)}00022\discretionary{%
}{-}{-}8}}


\bibitem{Sperrle2021MLEvaluations}
F.~Sperrle, M.~El-Assady, G.~Guo, R.~Borgo, D.~H. Chau, A.~Endert, and D.~Keim.
\newblock A survey of human-centered evaluations in human-centered machine
  learning.
\newblock {\em Comput. Graphics Forum}, 40(3):543--568, 2021.
  \href{https://doi.org/10.1111/cgf.14329}
{doi: \textsf{%
10\hspace{.1pt}\discretionary{.}{%
}{.}\hspace{.4pt}1111\discretionary{/}{%
}{/}cgf\hspace{.1pt}\discretionary{.}{%
}{.}\hspace{.4pt}14329}}


\bibitem{stolte2002polaris}
C.~Stolte, D.~Tang, and P.~Hanrahan.
\newblock Polaris: A system for query, analysis, and visualization of
  multidimensional relational databases.
\newblock {\em IEEE TVCG}, 8(1):52--65, 2002.
  \href{https://doi.org/10.1109/2945.981851}
{doi: \textsf{%
10\hspace{.1pt}\discretionary{.}{%
}{.}\hspace{.4pt}1109\discretionary{/}{%
}{/}2945\hspace{.1pt}\discretionary{.}{%
}{.}\hspace{.4pt}981851}}


\bibitem{Talukder2020Paletteviz}
A.~K.~A. Talukder and K.~Deb.
\newblock Paletteviz: A visualization method for functional understanding of
  high-dimensional pareto-optimal data-sets to aid multi-criteria decision
  making.
\newblock {\em IEEE CIM}, 15(2):36--48, 2020.
  \href{https://doi.org/10.1109/MCI.2020.2976184}
{doi: \textsf{%
10\hspace{.1pt}\discretionary{.}{%
}{.}\hspace{.4pt}1109\discretionary{/}{%
}{/}MCI\hspace{.1pt}\discretionary{.}{%
}{.}\hspace{.4pt}2020\hspace{.1pt}\discretionary{.}{%
}{.}\hspace{.4pt}2976184}}


\bibitem{Tian2021Projection}
Z.~Tian, X.~Zhai, G.~van Steenpaal, L.~Yu, E.~Dimara, M.~Espadoto, and
  A.~Telea.
\newblock Quantitative and qualitative comparison of 2d and 3d projection
  techniques for high-dimensional data.
\newblock {\em Information}, 12(6):239:1--239:21, 2021.
  \href{https://doi.org/10.3390/info12060239}
{doi: \textsf{%
10\hspace{.1pt}\discretionary{.}{%
}{.}\hspace{.4pt}3390\discretionary{/}{%
}{/}info12060239}}


\bibitem{tory2021finding}
M.~Tory, L.~Bartram, B.~Fiore-Gartland, and A.~Crisan.
\newblock Finding their data voice: Practices and challenges of dashboard
  users.
\newblock {\em IEEE CG\&A}, 43(1):22--36, 2023.
  \href{https://doi.org/10.1109/MCG.2021.3136545}
{doi: \textsf{%
10\hspace{.1pt}\discretionary{.}{%
}{.}\hspace{.4pt}1109\discretionary{/}{%
}{/}MCG\hspace{.1pt}\discretionary{.}{%
}{.}\hspace{.4pt}2021\hspace{.1pt}\discretionary{.}{%
}{.}\hspace{.4pt}3136545}}


\bibitem{tovanich2019visualization}
N.~Tovanich, N.~Heulot, J.-D. Fekete, and P.~Isenberg.
\newblock Visualization of blockchain data: a systematic review.
\newblock {\em IEEE TVCG}, 27(7):3135--3152, 2021.
  \href{https://doi.org/10.1109/TVCG.2019.2963018}
{doi: \textsf{%
10\hspace{.1pt}\discretionary{.}{%
}{.}\hspace{.4pt}1109\discretionary{/}{%
}{/}TVCG\hspace{.1pt}\discretionary{.}{%
}{.}\hspace{.4pt}2019\hspace{.1pt}\discretionary{.}{%
}{.}\hspace{.4pt}2963018}}


\bibitem{Pelt2014Blood}
R.~van Pelt, R.~Gasteiger, K.~Lawonn, M.~Meuschke, and B.~Preim.
\newblock Comparative blood flow visualization for cerebral aneurysm treatment
  assessment.
\newblock {\em Comput. Graphics Forum}, 33(3):131--140, 2014.
  \href{https://doi.org/10.1111/cgf.12369}
{doi: \textsf{%
10\hspace{.1pt}\discretionary{.}{%
}{.}\hspace{.4pt}1111\discretionary{/}{%
}{/}cgf\hspace{.1pt}\discretionary{.}{%
}{.}\hspace{.4pt}12369}}


\bibitem{Velasquez2013Analysis}
M.~Velasquez and P.~T. Hester.
\newblock An analysis of multi-criteria decision making methods.
\newblock {\em IJOR}, 10(2):56--66, 2013.

\bibitem{wall2017warning}
E.~Wall, L.~M. Blaha, L.~Franklin, and A.~Endert.
\newblock Warning, bias may occur: A proposed approach to detecting cognitive
  bias in interactive visual analytics.
\newblock In {\em Proc. VAST}, pp. 104--115. IEEE, Piscataway, 2017.
  \href{https://doi.org/10.1109/VAST.2017.8585669}
{doi: \textsf{%
10\hspace{.1pt}\discretionary{.}{%
}{.}\hspace{.4pt}1109\discretionary{/}{%
}{/}VAST\hspace{.1pt}\discretionary{.}{%
}{.}\hspace{.4pt}2017\hspace{.1pt}\discretionary{.}{%
}{.}\hspace{.4pt}8585669}}


\bibitem{Wall2018Podium}
E.~Wall, S.~Das, R.~Chawla, B.~Kalidindi, E.~T. Brown, and A.~Endert.
\newblock Podium: Ranking data using mixed-initiative visual analytics.
\newblock {\em IEEE TVCG}, 24(1):288--297, 2018.
  \href{https://doi.org/10.1109/TVCG.2017.2745078}
{doi: \textsf{%
10\hspace{.1pt}\discretionary{.}{%
}{.}\hspace{.4pt}1109\discretionary{/}{%
}{/}TVCG\hspace{.1pt}\discretionary{.}{%
}{.}\hspace{.4pt}2017\hspace{.1pt}\discretionary{.}{%
}{.}\hspace{.4pt}2745078}}


\bibitem{wang2022transparency}
R.~Wang, R.~Bush-Evans, E.~Arden-Close, E.~Bolat, J.~McAlaney, S.~Hodge,
  S.~Thomas, and K.~Phalp.
\newblock Transparency in persuasive technology, immersive technology, and
  online marketing: Facilitating users’ informed decision making and
  practical implications.
\newblock {\em Comput. Hum. Behav.}, 139:107545:1--107545:15, 2023.
  \href{https://doi.org/10.1016/j.chb.2022.107545}
{doi: \textsf{%
10\hspace{.1pt}\discretionary{.}{%
}{.}\hspace{.4pt}1016\discretionary{/}{%
}{/}j\hspace{.1pt}\discretionary{.}{%
}{.}\hspace{.4pt}chb\hspace{.1pt}\discretionary{.}{%
}{.}\hspace{.4pt}2022\hspace{.1pt}\discretionary{.}{%
}{.}\hspace{.4pt}107545}}


\bibitem{Waser2010WorldLines}
J.~Waser, R.~Fuchs, H.~Ribi{\v{c}}i{\v{c}}, B.~Schindler, G.~Bl{\"o}schl, and
  E.~Gr{\"o}ller.
\newblock World lines.
\newblock {\em IEEE TVCG}, 16(6):1458--1467, 2010.
  \href{https://doi.org/10.1109/TVCG.2010.223}
{doi: \textsf{%
10\hspace{.1pt}\discretionary{.}{%
}{.}\hspace{.4pt}1109\discretionary{/}{%
}{/}TVCG\hspace{.1pt}\discretionary{.}{%
}{.}\hspace{.4pt}2010\hspace{.1pt}\discretionary{.}{%
}{.}\hspace{.4pt}223}}


\bibitem{Waser2014ManyPlans}
J.~Waser, A.~Konev, B.~Sadransky, Z.~Horv{\'a}th, H.~Ribi{\v{c}}i{\'c},
  R.~Carnecky, P.~Kluding, and B.~Schindler.
\newblock Many plans: Multidimensional ensembles for visual decision support in
  flood management.
\newblock {\em Comput. Graphics Forum}, 33(3):281--290, 2014.
  \href{https://doi.org/10.1111/cgf.12384}
{doi: \textsf{%
10\hspace{.1pt}\discretionary{.}{%
}{.}\hspace{.4pt}1111\discretionary{/}{%
}{/}cgf\hspace{.1pt}\discretionary{.}{%
}{.}\hspace{.4pt}12384}}


\bibitem{Weistroffer2016MCDASoftware}
H.~R. Weistroffer and Y.~Li.
\newblock Multiple criteria decision analysis software.
\newblock In {\em Multiple criteria decision analysis: state of the art
  surveys}, chap.~29, pp. 1301--1341. Springer, 2016.
  \href{https://doi.org/10.1007/978-1-4939-3094-4_29}
{doi: \textsf{%
10\hspace{.1pt}\discretionary{.}{%
}{.}\hspace{.4pt}1007\discretionary{/}{%
}{/}978\discretionary{%
}{-}{-}1\discretionary{%
}{-}{-}4939\discretionary{%
}{-}{-}3094\discretionary{%
}{-}{-}4\_29}}


\bibitem{Weng2019SRVis}
D.~Weng, R.~Chen, Z.~Deng, F.~Wu, J.~Chen, and Y.~Wu.
\newblock Srvis: Towards better spatial integration in ranking visualization.
\newblock {\em IEEE TVCG}, 25(1):459--469, 2019.
  \href{https://doi.org/10.1109/TVCG.2018.2865126}
{doi: \textsf{%
10\hspace{.1pt}\discretionary{.}{%
}{.}\hspace{.4pt}1109\discretionary{/}{%
}{/}TVCG\hspace{.1pt}\discretionary{.}{%
}{.}\hspace{.4pt}2018\hspace{.1pt}\discretionary{.}{%
}{.}\hspace{.4pt}2865126}}


\bibitem{Weng2020Bus}
D.~Weng, C.~Zheng, Z.~Deng, M.~Ma, J.~Bao, Y.~Zheng, M.~Xu, and Y.~Wu.
\newblock Towards better bus networks: A visual analytics approach.
\newblock {\em IEEE TVCG}, 27(2):817--827, 2021.
  \href{https://doi.org/10.1109/TVCG.2020.3030458}
{doi: \textsf{%
10\hspace{.1pt}\discretionary{.}{%
}{.}\hspace{.4pt}1109\discretionary{/}{%
}{/}TVCG\hspace{.1pt}\discretionary{.}{%
}{.}\hspace{.4pt}2020\hspace{.1pt}\discretionary{.}{%
}{.}\hspace{.4pt}3030458}}


\bibitem{Weng2018Homefinder}
D.~Weng, H.~Zhu, J.~Bao, Y.~Zheng, and Y.~Wu.
\newblock Homefinder revisited: Finding ideal homes with reachability-centric
  multi-criteria decision making.
\newblock In {\em Proc. CHI}, pp. 247:1--247:12. ACM, New York, 2018.
  \href{https://doi.org/10.1145/3173574.3173821}
{doi: \textsf{%
10\hspace{.1pt}\discretionary{.}{%
}{.}\hspace{.4pt}1145\discretionary{/}{%
}{/}3173574\hspace{.1pt}\discretionary{.}{%
}{.}\hspace{.4pt}3173821}}


\bibitem{Wittenburg2001ParallelBargrams}
K.~Wittenburg, T.~Lanning, M.~Heinrichs, and M.~Stanton.
\newblock Parallel bargrams for consumer-based information exploration and
  choice.
\newblock In {\em Proc. UIST}, pp. 51--60. ACM, New York, 2001.
  \href{https://doi.org/10.1145/502348.502357}
{doi: \textsf{%
10\hspace{.1pt}\discretionary{.}{%
}{.}\hspace{.4pt}1145\discretionary{/}{%
}{/}502348\hspace{.1pt}\discretionary{.}{%
}{.}\hspace{.4pt}502357}}


\bibitem{xia2018dataink}
H.~Xia, N.~Henry~Riche, F.~Chevalier, B.~De~Araujo, and D.~Wigdor.
\newblock Dataink: Direct and creative data-oriented drawing.
\newblock In {\em Proc. CHI EA}, pp. D413:1--D413:13. ACM, New York, 2018.
  \href{https://doi.org/10.1145/3170427.3186471}
{doi: \textsf{%
10\hspace{.1pt}\discretionary{.}{%
}{.}\hspace{.4pt}1145\discretionary{/}{%
}{/}3170427\hspace{.1pt}\discretionary{.}{%
}{.}\hspace{.4pt}3186471}}


\bibitem{Yi2007Interaction}
J.~S. Yi, Y.~ah~Kang, J.~Stasko, and J.~A. Jacko.
\newblock Toward a deeper understanding of the role of interaction in
  information visualization.
\newblock {\em IEEE TVCG}, 13(6):1224--1231, 2007.
  \href{https://doi.org/10.1109/TVCG.2007.70515}
{doi: \textsf{%
10\hspace{.1pt}\discretionary{.}{%
}{.}\hspace{.4pt}1109\discretionary{/}{%
}{/}TVCG\hspace{.1pt}\discretionary{.}{%
}{.}\hspace{.4pt}2007\hspace{.1pt}\discretionary{.}{%
}{.}\hspace{.4pt}70515}}


\bibitem{Zavadskas2011MCDM}
E.~K. Zavadskas and Z.~Turskis.
\newblock Multiple criteria decision making (mcdm) methods in economics: an
  overview.
\newblock {\em Technol. Econ. Dev. Econ.}, 17(2):397--427, 2011.
  \href{https://doi.org/10.3846/20294913.2011.593291}
{doi: \textsf{%
10\hspace{.1pt}\discretionary{.}{%
}{.}\hspace{.4pt}3846\discretionary{/}{%
}{/}20294913\hspace{.1pt}\discretionary{.}{%
}{.}\hspace{.4pt}2011\hspace{.1pt}\discretionary{.}{%
}{.}\hspace{.4pt}593291}}


\bibitem{Zhang2015Remediating}
Y.~Zhang, R.~K. Bellamy, and W.~A. Kellogg.
\newblock Designing information for remediating cognitive biases in
  decision-making.
\newblock In {\em Proc. CHI}, pp. 2211--2220. ACM, New York, 2015.
  \href{https://doi.org/10.1145/2702123.2702239}
{doi: \textsf{%
10\hspace{.1pt}\discretionary{.}{%
}{.}\hspace{.4pt}1145\discretionary{/}{%
}{/}2702123\hspace{.1pt}\discretionary{.}{%
}{.}\hspace{.4pt}2702239}}


\bibitem{Zhao2018SkyLens}
X.~Zhao, Y.~Wu, W.~Cui, X.~Du, Y.~Chen, Y.~Wang, D.~L. Lee, and H.~Qu.
\newblock Skylens: Visual analysis of skyline on multi-dimensional data.
\newblock {\em IEEE TVCG}, 24(1):246--255, 2018.
  \href{https://doi.org/10.1109/TVCG.2017.2744738}
{doi: \textsf{%
10\hspace{.1pt}\discretionary{.}{%
}{.}\hspace{.4pt}1109\discretionary{/}{%
}{/}TVCG\hspace{.1pt}\discretionary{.}{%
}{.}\hspace{.4pt}2017\hspace{.1pt}\discretionary{.}{%
}{.}\hspace{.4pt}2744738}}


\bibitem{zhao2018evaluating}
Y.~Zhao, F.~Luo, M.~Chen, Y.~Wang, J.~Xia, F.~Zhou, Y.~Wang, Y.~Chen, and
  W.~Chen.
\newblock Evaluating multi-dimensional visualizations for understanding fuzzy
  clusters.
\newblock {\em IEEE TVCG}, 25(1):12--21, 2019.
  \href{https://doi.org/10.1109/TVCG.2018.2865020}
{doi: \textsf{%
10\hspace{.1pt}\discretionary{.}{%
}{.}\hspace{.4pt}1109\discretionary{/}{%
}{/}TVCG\hspace{.1pt}\discretionary{.}{%
}{.}\hspace{.4pt}2018\hspace{.1pt}\discretionary{.}{%
}{.}\hspace{.4pt}2865020}}


\bibitem{Ziegler2007Relevance}
H.~Ziegler, T.~Nietzschmann, and D.~A. Keim.
\newblock Relevance driven visualization of financial performance measures.
\newblock In {\em Proc. EuroVis}, pp. 19--26. Eurographics, Geneva, 2007.
  \href{https://doi.org/10.2312/VisSym/EuroVis07/019-026}
{doi: \textsf{%
10\hspace{.1pt}\discretionary{.}{%
}{.}\hspace{.4pt}2312\discretionary{/}{%
}{/}VisSym\discretionary{/}{%
}{/}EuroVis07\discretionary{/}{%
}{/}019\discretionary{%
}{-}{-}026}}


\end{thebibliography}

\end{document}